\documentclass[a4paper,twoside, 11pt]{report}
\usepackage[english]{babel}
\usepackage{amssymb}
\usepackage{amsmath}
\usepackage{graphicx}
\usepackage{acronym}
\usepackage{url}

\newcommand{\sd}{\mathrm{d}}

\newcommand{\isotope}[2]{\ensuremath{^{#1}\textrm{#2}}}

\newcommand{\ackgroup}[1]{\noindent#1\vspace{3.25pt}\begin{center}$\infty$\end{center}\vspace{3.25pt}}

\begin{document}
\pagestyle{empty}
\title{Development of Neutron Detectors for the Next Generation of Radioactive Ion-Beam Facilities} 
\author{Licentiate Thesis \\P\"{a}r-Anders S\"{o}derstr\"{o}m \\\\ 
\\\\\\\\\\\\ 
Department of Physics and Astronomy, Uppsala University} 
\date{June 2009}
\framebox[0.95\linewidth]{
\parbox[][][c]{0.95\linewidth}{
\maketitle}}

\cleardoublepage
\abstract{
The next generation of radioactive ion beam facilities, which will give
experimental access to many exotic nuclei, are presently being developed. These facilities will
make it possible to study very short lived exotic nuclei with extreme values of isospin far from
the line of ${\beta}$ stability. Such nuclei will be
produced with very low cross sections and to study them, new detector
arrays are being developed.

At the SPIRAL facility in GANIL a neutron detector array, the Neutron Wall, is
located. In this work the Neutron Wall has been characterized regarding neutron
detection efficiency and discrimination between neutrons and ${\gamma}$
rays. The possibility to increase the efficiency by increasing the high
voltage of the photomultiplier tubes has also been studied.

For SPIRAL2 a neutron detector array, NEDA, is being developed. NEDA
will operate in a high ${\gamma}$-ray background environment which puts
a high demand on the quality of discrimination between neutrons and
${\gamma}$ rays. To increase the quality of the discrimination methods pulse-shape discrimination techniques utilizing digital
electronics have been developed and evaluated regarding bit resolution and sampling frequency of the ADC. The
conclusion is that an ADC with a bit resolution of 12 bits and a
sampling frequency of 100 MS/s is adequate for pulse-shape
discrimination of neutrons and ${\gamma}$ rays for a neutron energy range of
0.3--12 MeV.
}
\vfill
\begin{tabular}{lr}
Supervisors: & J. Nyberg\\
 & A. Ata\c{c}\\
\end{tabular}
\cleardoublepage
\renewcommand{\thepage}{\roman{page}}
\setcounter{page}{1}
\pagestyle{plain}
\tableofcontents
\cleardoublepage
\chapter*{List of Acronyms}
\addcontentsline{toc}{chapter}{List of Acronyms}
\begin{acronym}
\acro{ADC}{analog-to-digital converter}
\acro{AGATA}{Advanced Gamma Tracking Array}
\acro{APD}{avalanche photodiode}
\acro{CFD}{constant-fraction discriminator}
\acro{DESCANT}{Deuterated Scintillator Array for Neutron Tagging}
\acro{DESPEC}{Decay Spectroscopy}
\acro{DSP}{digital signal processor}
\acro{EDEN}{Etude des D\'{e}croissances par Neutrons}
\acro{EURISOL}{European Isotope Separation On-Line}
\acro{FAIR}{Facility for Antiproton and Ion Research}
\acro{FPGA}{field programmable gate array}
\acro{FOM}{figure-of-merit}
\acro{FRS}{GSI Fragment Separator}
\acro{FWHM}{full width at half maximum}
\acro{GANIL}{Grand Accelerateur National d Ions Lourds}
\acro{GSI}{Gesellschaft f\"{u}r Schwerionenforschung GmbH}
\acro{HISPEC}{High-Resolution In-Flight Spectroscopy}
\acro{HPGe}{high-purity germanium}
\acro{HV}{high voltage}
\acro{IReS}{Institut de Recherches Subatomiques}
\acro{ISOL}{isotope separation on-line}
\acro{LED}{light emitting diode}
\acro{LINAC}{linear accelerator}
\acro{LNL}{Laboratori Nazionali di Legnaro}
\acro{MS/s}{megasamples per second}
\acro{NEDA}{Neutron Detector Array}
\acro{NIM}{Nuclear Instrumentation Module}
\acro{PC}{personal computer}
\acro{PMT}{photomultiplier tube}
\acro{PSA}{pulse-shape analysis}
\acro{PSD}{pulse-shape discrimination}
\acro{QVC}{charge-to-voltage converter}
\acro{RIB}{radioactive ion beam}
\acro{SPIRAL}{Syst\`{e}me de Production d'Ions Radioactifs en Ligne}
\acro{Super-FRS}{Super-conducting Fragment Separator}
\acro{TAC}{time-to-amplitude converter}
\acro{TNT}{Tracking Numerical Treatment}
\acro{TOF}{time-of-flight}
\acro{ZCO}{zero cross-over}
\end{acronym}
\cleardoublepage
\addcontentsline{toc}{chapter}{List of Figures}
\listoffigures
\cleardoublepage

\chapter{Preface}
\renewcommand{\thepage}{\arabic{page}}
\setcounter{page}{1}
\label{intro}
This licentiate thesis is based on the paper {\em Digital pulse-shape discrimination of fast neutrons and $\gamma$ rays} by P.-A. S\"{o}derstr\"{o}m, J.~Nyberg, and R.~Wolters \cite{dpsd} (reprinted with permission from Elsevier). It concerns the detection of fast neutrons for applications in $\gamma$-ray spectroscopy of exotic nuclei using organic liquid scintillator detector arrays. Since I have an interest in the the history of science, I  find the neutron very special. It took only ten years from its discovery in 1932 \cite{mabyexistneutron} until it in 1942 had found its application in the artificial nuclear reactor under the stands of a football stadium at the University of Chicago \cite{fermis_pile}. Physics involving fast neutrons also lies close to me since neutron induced nuclear reactions \cite{p-a_exjobb,nd2007_hayashi,nd2007_tippawan} was what got me involved in experimental nuclear physic in the first place.

 My impression is that there is a large gap in the level of information between the standard detector reference book by Knoll \cite{knoll}, which gives an introduction to the topic, and high level scientific papers. One aim of this thesis is to try to partially fill this gap in order for future students to faster get acquainted with the topic. Therefore, the layout of the thesis is to first give an introduction to the physics case in chapter~\ref{sec:exoticns}, followed by an introduction to liquid scintillators and how these are used to detect fast neutrons in chapter~\ref{sec:detection}. The Neutron Wall detector array and the results from a characterization test are presented in chapter~\ref{sec:NWall}, where I have contributed to the results in section~\ref{sec:status}. Chapter~\ref{sec:dpsd} concerns my work on  digital pulse-shape analysis and is focused on ref.~\cite{dpsd}. Finally, in chapter~\ref{sec:neda} an outlook is given towards the next generation neutron detector array NEDA that is being developed as a part of the SPIRAL2 Preparatory Phase project, where my contribution is evaluating possible readout systems presented in section~\ref{ss:readout}. The name NEDA comes from the Okeanid Naiad Nymph of the river Neda near Mount Lykaios in the southern Greek region Arkadia.
\cleardoublepage

\chapter{Nuclear Structure Far From Stability\label{sec:exoticns}}
\begin{figure}
 \centering
 \includegraphics[width=\textwidth]{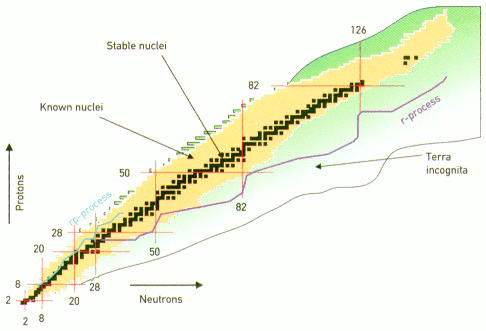}
 \caption[The Segr\'{e} chart of nuclear isotopes]{The Segr\'{e} chart of nuclear isotopes. Courtesy of Witek Nazarewicz.}
 \label{fig:nchart}
\end{figure}

Since a long time it is well known that matter is made up of atoms. The atoms in turn are composed of an atomic nucleus surrounded by a cloud of electrons. This nucleus is the subject that is studied in the field of nuclear structure physics. A nuclide is an atomic nucleus with a specific number of protons and neutrons. For lighter nuclei the number of protons and neutrons are approximately the same, while heavier nuclei consist of more neutrons than protons. The chart of nuclides, or the Segr\'{e} chart, is a plot of the number of protons versus the number of neutrons, see fig.~\ref{fig:nchart}.

As seen in the Segr\'{e} chart, the number of stable nuclei are very few. Arranged in a bent line called the line of $\beta$ stability, there are only about 250 of them. Many more nuclei can however be constructed either in laboratories on earth, or in violent astrophysical events like supernovae explosions. About 3000 elements have up to now been created and observed in laboratories, but theorists predict that more than 6000 bound nuclei can exist between the neutron and proton drip-lines, which are defined as the limits of existence. The physics of these very exotic nuclides is to a large extent unknown and history has thought us that many surprises probably await in this {\em terra incognita}.

Now, just around the corner, awaits the second generation \ac{RIB} facilities which will give the possibility to travel far into the unknown parts of the Segr\'{e} chart. Such facilities are for example the Super-FRS at FAIR \cite{sfrs,2007JPhG...34..551H} and SPIRAL2 \cite{2007PrPNP..59...22G} that will evolve into EURISOL \cite{eurisol}. This is sometimes referred to as the fourth revolution within nuclear spectroscopy \cite{riccirecollection}. The first revolution was the discovery of NaI(Tl) scintillator detectors, by which one could start to measure quite accurately the energy and intensity of $\gamma$-ray transitions in various radioactive nuclei. These measurements later radically improved when one started using semiconductor detectors of germanium instead of NaI(Tl) crystals. The second revolution came with the possibilities to use in-beam $\gamma$-ray spectroscopy together with heavy-ion nuclear reactions. This made it possible to study specific levels of the nuclei of interest. The latest revolution is the development of sophisticated high-resolution spectroscopy arrays of high-purity germanium, crowned by EUROBALL \cite{simpseuro} and GAMMASPHERE \cite{gammasphere}. These gave access to very weak signals from high-spin states, unraveling many new nuclear structure phenomena. A few frontiers in the Segr\'{e} chart to be studied at the \ac{RIB} facilities, will be briefly discussed in this chapter\footnote{This is just a small selection of the physics case for the radioactive ion beam facilities and is by no means intended to be complete but rather a reflection of the authors personal favorite topics.}.

\section{Frontiers in Nuclear Structure}

\subsection{Light Neutron-rich Nuclei}
\label{sec:lightn}

The only area in the Segr\'{e} chart where the neutron drip-line has been reached is the area containing the light neutron-rich nuclei. This area is also an example on how unexpected phenomena can appear when moving away from the line of $\beta$ stability. The first time this area was explored was using high-energy fragmentation reactions and the results were that some neutron-rich nuclei had abnormally large reaction cross-sections, in particular $^{11}$Li. This was interpreted as a result of the size of these nuclei being much larger than expected. In fact, $^{11}$Li has the same size as $^{208}$Pb despite the much fewer number of nucleons. To understand this, a model was suggested that $^{11}$Li actually consists of a $^{9}$Li core surrounded by a halo of two neutrons, a picture that was confirmed by measurements of the quadrupole moment \cite{Arnold199216} and the charge radius \cite{sanchez:033002} of $^{11}$Li.

With the new intense \ac{RIB}s there are good opportunities for studying $^{11}$Li and other Borromean\footnote{This name comes from the Borromean family crest which is made of three rings intwined such that if one is removed the entire system falls apart. In the same way the Borromean halo nuclei consist of a nucleus and a halo of two neutrons, while neither system of the specific nucleus and one neutron halo nor the system of two neutrons are bound.} halo nuclei more closely. The Neutron Wall, to be described in a later chapter, has already been used together with an \ac{ISOL} beam to study neutron configurations in this kind of nuclei \cite{chatterjee:032701}. For more information see for example \cite{Zhukov1993151}.

\subsection{Heavy Proton-rich Nuclei}
\label{sec:heavyp}

To test the nuclear shell model it is of much interest to explore the proton-rich side of the line of $\beta$ stability. One example of a proton rich exotic nucleus is \isotope{100}{Sn}, the heaviest self-conjugate doubly-magic nucleus that is expected to be bound. This itself is an interesting property but apart from this it represents a unique testing ground for various aspects of the shell model, like single-particle energies by examining the neighbors with one \cite{seweryniak:022504,sn101npn} or two particles or holes outside the closed shell \cite{1996ZP,magdas}. Since the neutrons and protons occupy the same orbits there is also a good opportunity to test the effects of neutron-proton pairing.

\begin{figure}
 \centering
 \includegraphics[width=\textwidth]{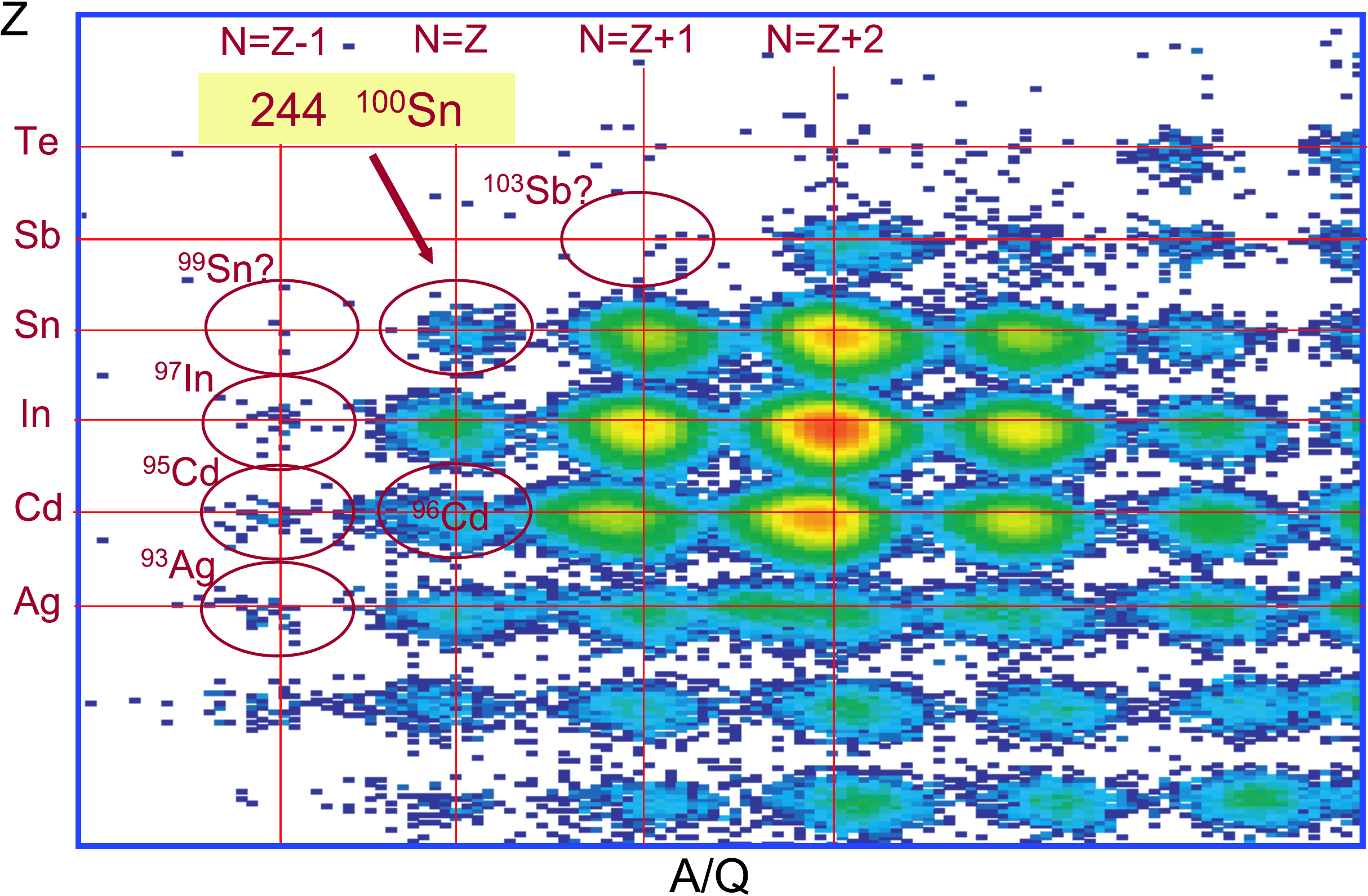}
 \caption[Identification of nuclei in the $^{100}$Sn neighborhood]{Identification of nuclei in the $^{100}$Sn neighborhood using fragmentation of $^{124}$Xe at GSI. Figure from \cite{gsi_sn100}.}
 \label{fig:sn100_gsi}
\end{figure}

The isotope $\isotope{100}{Sn}$ was discovered by two independent experiments \cite{sn100_94,sn100_94_2} but the number of events was very low in both experiments. Recently it has been shown that it is possible to produce about one $\isotope{100}{Sn}$ per hour using fragmentation reactions at the FRS at GSI \cite{gsi_sn100}, see fig.~\ref{fig:sn100_gsi}. This rate should be dramatically increased at FAIR and also become available for fusion-evaporation reactions at the next \ac{ISOL} facility SPIRAL2 at GANIL \cite{2007PrPNP..59...22G}.

\subsection{Heavy Neutron-rich Nuclei}
\label{sec:heavyn}

While the study of exotic doubly-magic nuclei is an interesting topic there are also so-called doubly mid-shell nuclei which are even more rare. Only four even-even doubly mid-shell nuclei are expected to be particle bound: \isotope{28}{Si}, \isotope{38}{Si}, \isotope{48}{Cr} and \isotope{170}{Dy}. 

Analogous to \isotope{100}{Sn} being a unique testing ground for the shell model, \isotope{170}{Dy} should also be a unique for testing of the evolution of collectivity. Due to its doubly mid-shell nature one can naively assume it to be the most collective of all nuclei. Calculations also suggest a stiff prolate axially symmetric shape \cite{paddy}, something that makes it interesting in terms of how stiffness of quadrupole deformation would delay backbending, a phenomenon that was first observed in \isotope{160}{Dy} \cite{1971PhLB...34..605J}. Nuclei in this area are further known to have states with large single-particle angular momentum components along the symmetry axis, $K$, that decay to the ground state band with $K=0$ \cite{highK}. This gives rise to long-lived isomeric states and \isotope{170}{Dy} is a promising candidate to have a very pure high-$K$ isomer \cite{paddy}. It is also the best candidate in the entire Segr\'{e} chart for empirical realization of the dynamical SU(3) symmetry in the interacting boson model \cite{ibm,PhysRevC.31.1991}.

This is an area that is very difficult to reach using today's facilities. The most successful attempt so far is using multi-nucleon transfer reactions \cite{170dy_lnl}. With future \ac{RIB} facilities, the accessibility is expected to increase dramatically.

\subsection{Super-heavy Nuclei}
\label{sec:sheavy}

Besides testing how far one can reach on the proton- and neutron-rich sides of the Segr\'{e} chart, it is also of much interest to see how far one can follow the $\beta$-stability line to produce as heavy elements as possible. This is not only of interest from a nuclear physics point of view, but also because this means that one would produce completely new chemical elements. Looking at fig.~\ref{fig:nchart} it seems there is not much beyond uranium and the actinides to study. However, it is suggested that there should exist additional higher magic numbers than those shown in fig.~\ref{fig:nchart}. According to theoretical predictions these new magic numbers would be $N=184$ and $Z=114$ or $Z=126$, which could cause an ``Island of Stability'' with long-lived elements to appear. For further details regarding the discovery of super-heavy elements see \cite{superheavydisc} and references therein.

One of the problems with the creation of super-heavy elements produced in fusion-evaporation reactions with stable beams is that one cannot reach the $N=184$ shell gap. Instead, it is suggested to use intense radioactive neutron-rich beams, which may dramatically increase some of the cross sections for production of super-heavy elements \cite{superheavyrib}.

\section{Radioactive Ion Beam Facilities}

To create \ac{RIB}s, there are two techniques that are in principle complementary to each other: the in-flight fragmentation and the \ac{ISOL} technique. 

In the in-flight fragmentation technique, a heavy-ion beam is accelerated to high energies, typically to about 50\% of the speed of light. The heavy-ion beam is chosen to be either close to the desired isotope or a fissile isotope like uranium. To produce the \ac{RIB}, one lets the primary heavy ion beam hit a light target like beryllium after which it will either fragment or fission into a cocktail of reaction products. The reaction products are then separated by a system of dipole magnets, quadrupole magnets and energy degraders. This technique is implemented at, for example, the FRS at GSI \cite{frs}. Using this technique one can produce exotic nuclides at relativistic energies which are limited only by the reaction cross-section and the flight time through the magnetic spectrometer, which is typically in the order of 100~ns.

The other method is the \ac{ISOL} technique, which to a large extent is the complete opposite to the in-flight fragmentation method. In an \ac{ISOL} facility one typically uses a high-intensity proton, neutron, deuteron or even electron beam on a heavy target, for example UC$_{x}$. The energy of the beam only has be high enough to give a high fission yield in the target, in which the fission fragments are completely stopped and neutralized. The fission fragments are extracted from the target chemically through diffusion and effusion into an ion source that ionizes the fragments. A post accelerator is then used to accelerate them to the desired energies. The limitations of this method is mainly due to the chemical extraction properties of the desired beam which can vary largely between different elements. See \cite{isol} for more details regarding the \ac{ISOL} technique.

\subsection{HISPEC/DESPEC at FAIR}

The next generation of in-flight fragmentation \ac{RIB} facility that will be built is the Super-FRS \cite{sfrs} at FAIR. The Super-FRS will consist of two achromatic degrader systems and superconducting magnets. It will also separate fission fragments, something that the current FRS can not do. Two of the experimental setups that will be coupled to the Super-FRS are HISPEC \cite{2006IJMPE..15.1967P} and DESPEC \cite{2006IJMPE..15.1979R}.

HISPEC is a detector system for in-flight spectroscopy of reactions using \ac{RIB}s. With this setup one can perform $\gamma$-ray, charged particle and neutron spectroscopic measurements using Coulomb excitation and fragmentation reactions. High-resolution \ac{HPGe} detector arrays, one of which will be the AGATA spectrometer \cite{Simpson_npn13-4,agata_pos,agata_test}, will be used together with a particle identification system after the secondary target.

The other experimental setup, which will be placed further down the same beam line as HISPEC, is DESPEC. DESPEC will be a setup for studying the decay properties of exotic nuclei. This will be done by using a highly pixellated implantation detector to stop the high energy \ac{RIB}, a dedicated high resolution \ac{HPGe} detector array and a high-efficiency neutron detector array. The plan is to design this neutron detector using the liquid scintillator BC501A and to identify the particle and measure its energy using a combination of \ac{PSA} in the neutron detector and \ac{TOF} between the implantation and the neutron detectors.

\subsection{EXOGAM2 at GANIL\label{sec:spiral2}}

The next generation \ac{ISOL} \ac{RIB} facility is planned to be EURISOL. Since EURISOL is further in the future than FAIR, there is an intermediate-generation facility planned at GANIL, called SPIRAL2. The SPIRAL2 facility will be based on a high power superconducting driver \ac{LINAC} that will deliver a 40~MeV deuteron beam with a beam current of 5~mA. This beam will be directed on a UC$_x$ target that produces the isotopes for the secondary beam. The driver \ac{LINAC} will also be able to produce very intense stable beams to be used either directly in experiments or for creation of proton-rich \ac{RIB}s via fusion-evaporation reactions. For this new facility the existing HPGe array at GANIL, EXOGAM \cite{exogam}, will be upgraded to EXOGAM2. EXOGAM2 will use digital electronics and be able to run in triggerless mode and it may be used together with AGATA.

In the SPIRAL2 preparatory phase work package 5, there is a specific task (task~5.8) to construct a neutron detector array to be used together with EXOGAM2. Below is a quotation from an internal document within task~5.8 about the requirements of this neutron detector array, NEDA:

\begin{quotation}
 The Neutron Detector will be built as an ancillary detector for a germanium array installed at the intense stable and radioactive ion beams from SPIRAL2. The Neutron Detector should be characterized by the highest possible neutron detection efficiency, excellent discrimination of neutrons and gamma rays, and a very small neutron-scattering probability. These properties are necessary in order to achieve a clean and efficient identification of gamma rays from the rare neutron-deficient nuclides produced in reactions in which only two or more neutrons are emitted.
\end{quotation} 

This licentiate thesis concerns the development of electronics and readout for NEDA.
\cleardoublepage

\chapter{Neutron Detection with Liquid Scintillators\label{sec:detection}}
Due to their uncharged nature, detection of neutrons is much more difficult than detection of charged particles. Yet, there are a variety of ways to detect fast neutrons. Some examples are to use elastic scattering off protons in a scintillator, charged particle reactions on $^{3}$He in ionization chambers, radiative capture on gadolinium, fission of a uranium foil on a silicon wafer. For more possible detection methods see ref.~\cite{knoll}. This chapter presents an introduction to liquid scintillators and how these are used to detect fast neutrons. The fast-neutron detection method focused on here is to use liquid organic scintillators. This type of scintillator mainly consist of aromatic carbohydrates, with small admixtures of other molecules. The neutrons are detected mainly due to elastic scattering with protons.

\section{Scintillation Process in Organic Materials}

\begin{figure}
 \centering
 \includegraphics[width=0.75\textwidth]{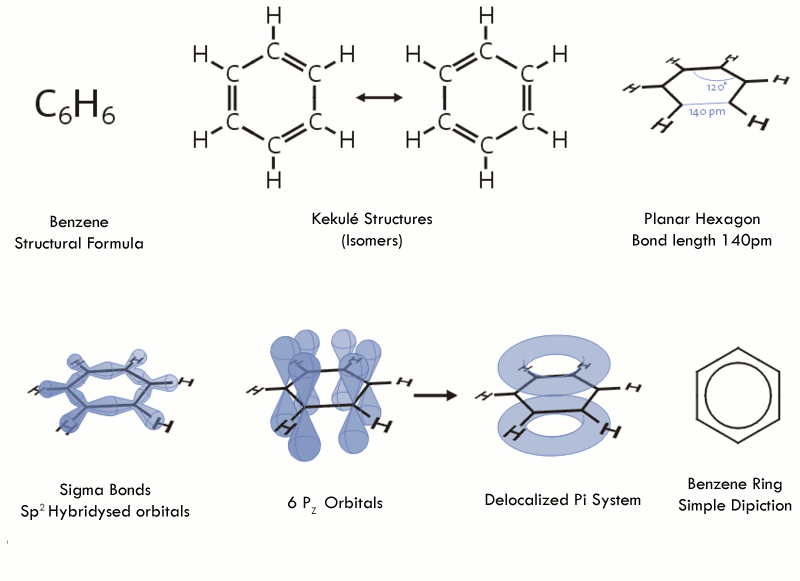}
 \caption[Binding mechanism of benzene]{Binding mechanism of benzene \cite{benzene}.}
 \label{fig:benzene}
\end{figure}

In aromatic carbohydrates there are two types of chemical covalent bonds that are important, the $\sigma$ bond and the $\pi$ bond. A carbon atom ready for binding will have an electron configuration of $1s^{2}2s2p^{3}$, meaning one valence electron in an $s$ orbital and three valence electrons in $p$ orbitals. Since the $s$ electron orbital is spherically symmetric it will always form axially symmetric $\sigma$ bonds while the $p$ electron orbitals (orbitals with two lobes) can form either axially symmetric $\sigma$ bonds or mirror symmetric $\pi$ bonds. The $\sigma$ bonds are the normal regular tetrahedron bonds of carbon and they do not contribute to the luminescence of the liquid, whereas the $\pi$ bonds cause double and triple bonds and are responsible for the luminescence. In the aromatic carbohydrates there are several of these $p$ orbitals that make up a delocalized $\pi$ system that can be modeled as free electrons orbiting the molecule. See fig.~\ref{fig:benzene} for an example of the benzene molecule binding and refs.~\cite{OrganicChemistry,birks} for more details on the structure of organic molecules.

\begin{figure}
 \centering
 \includegraphics[width=0.60\textwidth]{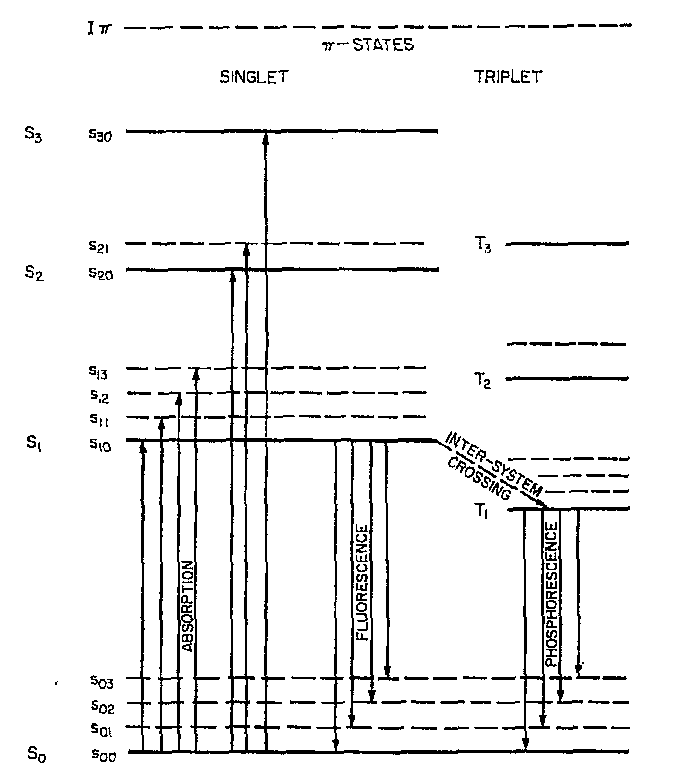}
 \caption[Typical level scheme of an organic scintillator]{Typical level scheme of an organic scintillator. Figure from \cite{birks}.}
 \label{fig:scintlevels}
\end{figure}

From the free electron model of the $\pi$ electrons ($p$ electrons in a delocalized $\pi$ system) moving in a one-dimensional circle, the $\pi$ electron will have a certain ground state and a number of excited singlet states (see fig.~\ref{fig:scintlevels}). These excited states will have an increasing excitation energy up to the ionization energy $I_\pi$, corresponding to different modes of the wave function. These states are usually denoted $S_0$ for the ground state and $S_1, S_2, \ldots$ for the excited states. The molecule can also be excited into triplet states denoted $T_{1}, T_{2}, \ldots$ that are usually lower in energy than their singlet counterparts. These states are further divided into vibrational sub-states as $S_{00}, S_{01}, \ldots, S_{10}, S_{11}, \ldots$ and $T_{10}, T_{11}, \ldots, T_{20}, T_{21}, \ldots$.
This kind of structure gives rise to three types of light that can be produced in the liquid scintillator: fluorescence, phosphorescence and delayed fluorescence\footnote{This is the standard terminology. For the rest of the text the mention of scintillation light, fast component and slow component will refer to the fluorescence and its time structure if not explicitly stated otherwise.}. The fastest of these processes is the fluorescence where the molecule de-excites through a series of $S$ states. The fluorescent light will only occur from transitions of type $S_{10} \to S_{0x}$ and has a lifetime of $\sim1$--$10$~ns. The molecule can also undergo a non-radiative transition from the $S_1$ state to some metastable state $M$, which in some cases been has identified as $T_1$ \cite{LewisKasha}, but could be any metastable state. From $M$ the molecule can de-excite to $S_0$ emitting a phosphorescence photon with a different energy than the fluorescence light and a lifetime $\gtrsim100$~$\mu$s. Another possible decay path from $M$ is that the molecule is thermally excited back to $S_1$ which then de-excites to $S_0$ emitting delayed fluorescence with an increased lifetime of $\sim1$~$\mu$s. In this work the main interest is in the fluorescence.

In some scintillators the fluorescence shows a time structure that is not described by a simple exponential decay, but which rather is a superposition of several exponential decay components with different decay time constants. All these components show the same spectrum as the fluorescence, that is the $S_{10} \to S_{0x}$ transition. It has also been observed that the slower component is largely unaffected by the decrease in fluorescence intensity, quenching, when particles with different ionization and excitation density are detected \cite{ParkerHatchard}. It has been proposed that the slower component is due to the interaction of two triplet-excited molecules that produce a transient dimer (two similar but separate molecules which are held together by intramolecular or intermolecular forces) which subsequently decomposes to give a molecule in the $S_1$ excited state and a molecule in the $S_0$ ground state \cite{knoll}.

\begin{figure}
 \centering
 \includegraphics[width=\textwidth]{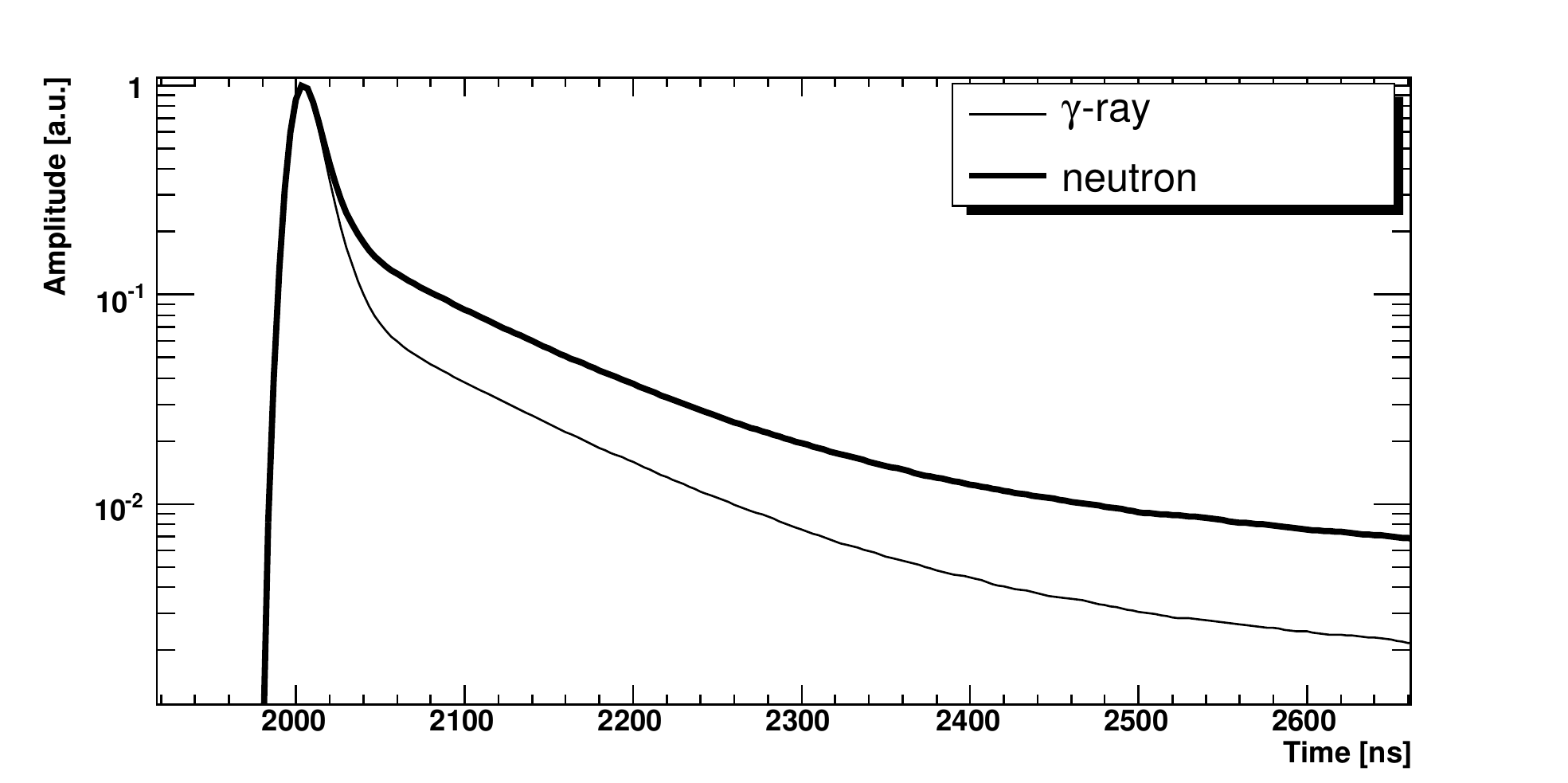}
 \caption[Idealized pulse shapes from a BC-501 liquid scintillator from a $\gamma$ ray and a neutron interaction]{Idealized pulse shapes from a BC501 liquid scintillator from a $\gamma$-ray and a neutron interaction. The decay times are 3.16~ns, 32.3~ns and 270~ns. Figure from \cite{varenna_pa}.\label{fig:ngamma}}
\end{figure}

That quenching does not significantly effect the slow florescence component is the cause of the pulse-shape difference for different particle species, see fig.~\ref{fig:ngamma}.
This means that although an $\alpha$ particle interaction gives rise to a relatively larger slow component than an electron interaction, the total number of photoelectrons are not necessarily larger for the $\alpha$ particle.
The relative intensity of the slow component for different particle species was first measured by Owen \cite{Owen1959a}\footnote{This is often erroneously cited as \cite{Owen1959b}. This is a misunderstanding that probably origins from ref.~\cite{birks} where the correct citation is used and these two references are denoted Owen 1959a and Owen 1959b respectively. Ref.~\cite{Owen1959b}, or Owen 1959b, however only concerns the observation of the slow scintillation component and not a measurement of it.}. Refined measurements were made by Gibbons et al. \cite{gibbonstable}, which are shown as an example in table~\ref{tab:taildependence}.
\begin{table} [ht] 
  \caption[Dependence of tail emission in anthracene on the species of exciting particle]{Dependence of tail emission in anthracene on the species of exciting particle\label{tab:taildependence}. Adapted from ref.~\cite{gibbonstable}.}
  \begin{tabular*}{\columnwidth}{@{\extracolsep{\fill}}cccc}
    \hline
     & Rel. intensity & Tail emission & Delayed photons\\
     & (main pulse) & (\% of main pulse) & per MeV \\
    \hline
    5 MeV $\alpha$ particle &  1  &   20    &  350    \\
    5 MeV neutron &  4  &    14  &    980  \\
    1.1 MeV $\gamma$ ray &  10  &    3.5   &   610     \\
    \hline
  \end{tabular*}
\end{table}

\section{The BC-501A liquid scintillator}

 One very popular liquid scintillator for neutron detection is BC-501A\footnote{In older detector arrays a similar liquid called BC-501 was used. These are manufactured by Saint-Gobain Ceramics \& Plastics. There is also an equivalent liquid from Nuclear Enterprise known as NE-213}. This is a scintillator liquid based on xylene or dimethylbenzene, C$_{6}$H$_{4}$(CH$_{3}$)$_{2}$. Xylene is flammable with a flash point (the temperature where it can form an ignitable mixture in air) of 24 degrees and poisonous since it can cause neurological damage at high exposures.

The liquid BC-501A has a light output that is about 78\% of anthracene, a maximum emission wavelength of 425~nm and a hydrogen to carbon ratio of 1.287. It has three decay components with 3.16~ns, 32.3~ns and 270~ns decay times \cite{KuchnirLynch}. The 270~ns component is mainly responsible for the \ac{PSD} properties.

\section{Detector Units\label{ss:detectors}}

\begin{figure}
 \centering
 \includegraphics[width=\textwidth]{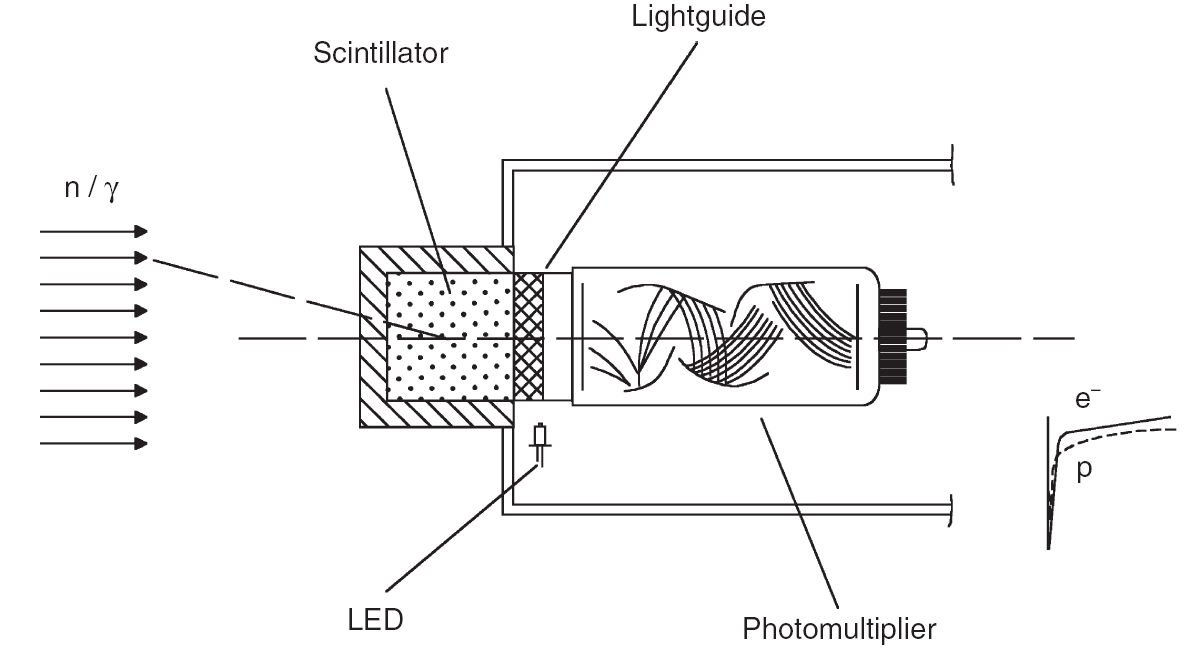}
 \caption[Process of neutron detection in a liquid scintillator]{Process of neutron detection in a liquid scintillator. Reprinted from \cite{scintfig} with permission from Elsevier.}
 \label{fig:scintdet}
\end{figure}

Beside the liquid, a neutron detector unit also consists of a \ac{PMT} for readout. See fig.~\ref{fig:scintdet} for an illustration of a typical neutron detector unit. A \ac{PMT} is a read out device that consist of a photocathode from which the incoming scintillation photons emits photoelectrons through the photo-electric effect. These photoelectrons are then accelerated in a \ac{HV} electric field towards a dynode, at which several more electrons are emitted. This process is repeated a number of times and the resulting electric current is read out from the anode at the end of the \ac{PMT}. The \ac{PMT} is coupled to the scintillator via a light guide, for example a glass window. Sometimes a \ac{LED} is also included in the detector unit for testing purposes.

For nuclear physics experiments this kind of detector units are usually organized into large arrays of different shapes and sizes, see refs.~\cite{1991NIMPA.300..303A,eden,Tilquin1995446,1999NIMPA.421..531S,2004NIMPA.530..473S} for a small selection of examples.

One important property of a liquid scintillator detector is the ability to discriminate between the detection of $\gamma$ rays and neutrons. This is not only important in order to get a clean neutron signal, but it has also been shown that even a small amount of $\gamma$ rays misinterpreted as neutrons dramatically reduce the quality of the cross-talk rejection \cite{2004NIMPA.528..741}.

\section{Discrimination Between Neutrons and $\gamma$ rays\label{ss:disc}}

Several different \ac{PSA} methods are used for discriminating between neutrons and $\gamma$ rays. Two well known methods are the \ac{ZCO} method, where the signal is shaped into a bipolar pulse from which the zero crossing is extracted \cite{alexander1961,roush1964psd}, and the charge comparison method, where the charge in the fast component is integrated and compared to the charge in the slow component \cite{brooks}. Both these methods give a cleaner separation for high-energy neutrons compared to low-energy neutrons since the statistics in the slow component of the pulse is higher for high-energy interactions.

\begin{figure}
 \centering
 \includegraphics[width=0.8\textwidth]{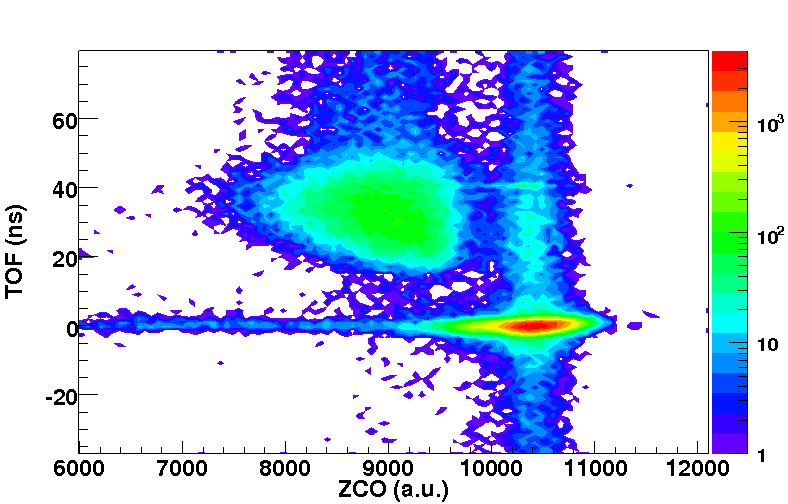}
 \caption[Example of a two-dimensional n-$\gamma$ spectrum]{Example of a two-dimensional n-$\gamma$ spectrum with the time-of-flight spectrum on the $y$ axis and the zero cross-over spectrum on the $x$ axis, obtained from the data in \cite{dpsd}. The upper left distribution is the neutron peak and the lower right distribution is the $\gamma$-ray peak. The vertical and horizontal $\gamma$-ray distributions are random coincidences and pile-up events, respectively.\label{fig:ng2d}}
\end{figure}

Another independent method to discriminate against different interacting particles is by measuring the \ac{TOF} between a global trigger and the interaction time of the signal in the detector. Since $\gamma$ rays travel with the speed of light, $c$, while the neutrons have a finite velocity, $v<c$, the  $\gamma$ rays will end up in a narrow peak in a \ac{TOF} spectrum while the neutrons will give rise to a distribution which depends on the neutron energy distribution. Since high-energy neutrons have a velocity closer to $c$ than low energy neutrons this discrimination technique works best for low-energy neutrons. Due to the independent and complementary nature of these discrimination techniques both of them are usually used together and the selection of neutrons is made in a two-dimensional spectrum, see fig.~\ref{fig:ng2d}.
\cleardoublepage

\chapter{The Neutron Wall\label{sec:NWall}}
The Neutron Wall detector array is presented together with the results from a characterization test performed in 2009 (section~\ref{sec:status}) in this chapter .

\section{Overview of the Neutron Wall}

\begin{figure}
 \centering
 \includegraphics[width=\textwidth]{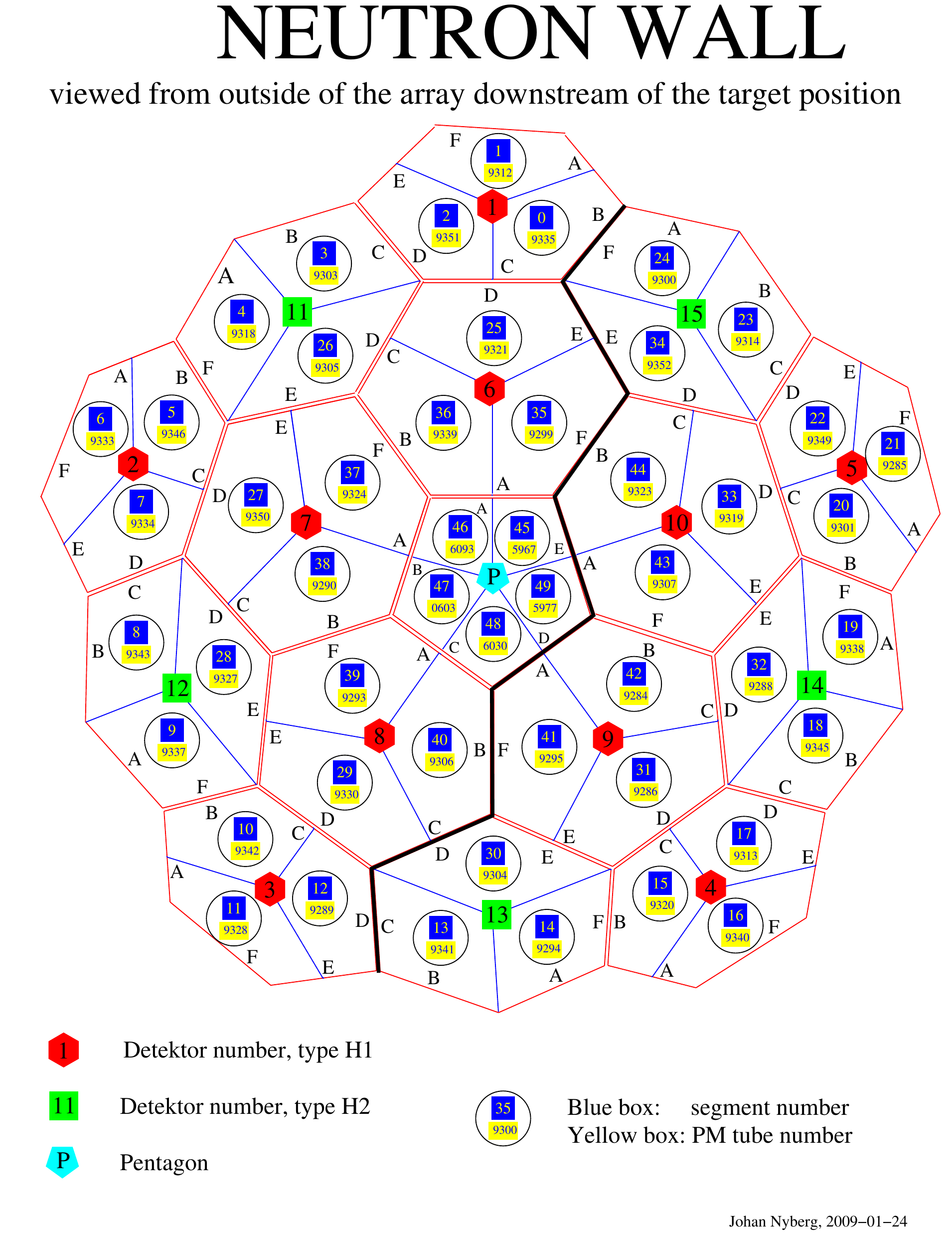}
 \caption[Sketch of the Neutron Wall]{Sketch of the Neutron Wall with the numbering of detectors, segment and PMTs.}
 \label{fig:nwall_numbers}
\end{figure}

An existing and so far very successful, neutron detector array for use in nuclear structure experiments is the Neutron Wall \cite{1999NIMPA.421..531S}. The Neutron Wall was originally designed for experiments together with EUROBALL \cite{simpseuro} at LNL and IReS Strasbourg, but since 2005 it is located at GANIL where it is used together with EXOGAM \cite{exogam}. It consists of 15 hexagonal detectors of two different shapes and one pentagonal detector. The detectors are assembled into a closely packed array covering about $1\pi$  of the solid angle. They are filled with the liquid scintillator BC-501A to a total volume of 150 litre. The 16 detectors are in turn divided into 50 segments as shown in fig.~\ref{fig:nwall_numbers}.
The hexagonal detectors are subdivided into three individual segments. Each of the segments contain 3.23 litre of scintillation liquid and is read out by a 130~mm Philips XP4512PA \ac{PMT}. The pentagonal detector is subdivided into five individual segments of 1.07 litre each and is read out by a 75~mm Philips XP4312B \ac{PMT}s.
 The \ac{PMT}s cover an average of 47\% and 50\% of the exit face of the hexagonal and pentagonal detectors, respectively. The total neutron efficiency of the Neutron Wall is about 25\% in symmetrical fusion-evaporation reactions.

\subsection{PSD Electronics\label{ss:psd}}

\begin{figure}
 \centering
 \includegraphics[width=\textwidth]{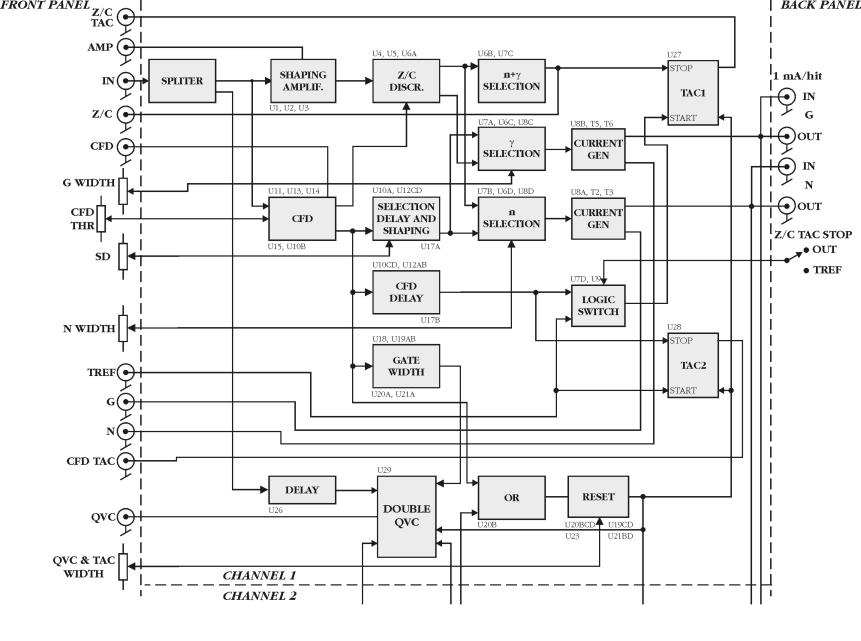}
 \caption[NDE202 block diagram]{NDE202 block diagram of one channel. Figure from \cite{NDE202_manual}.}
 \label{fig:nde202block}
\end{figure}

\begin{figure}
 \centering
 \includegraphics[width=\textwidth]{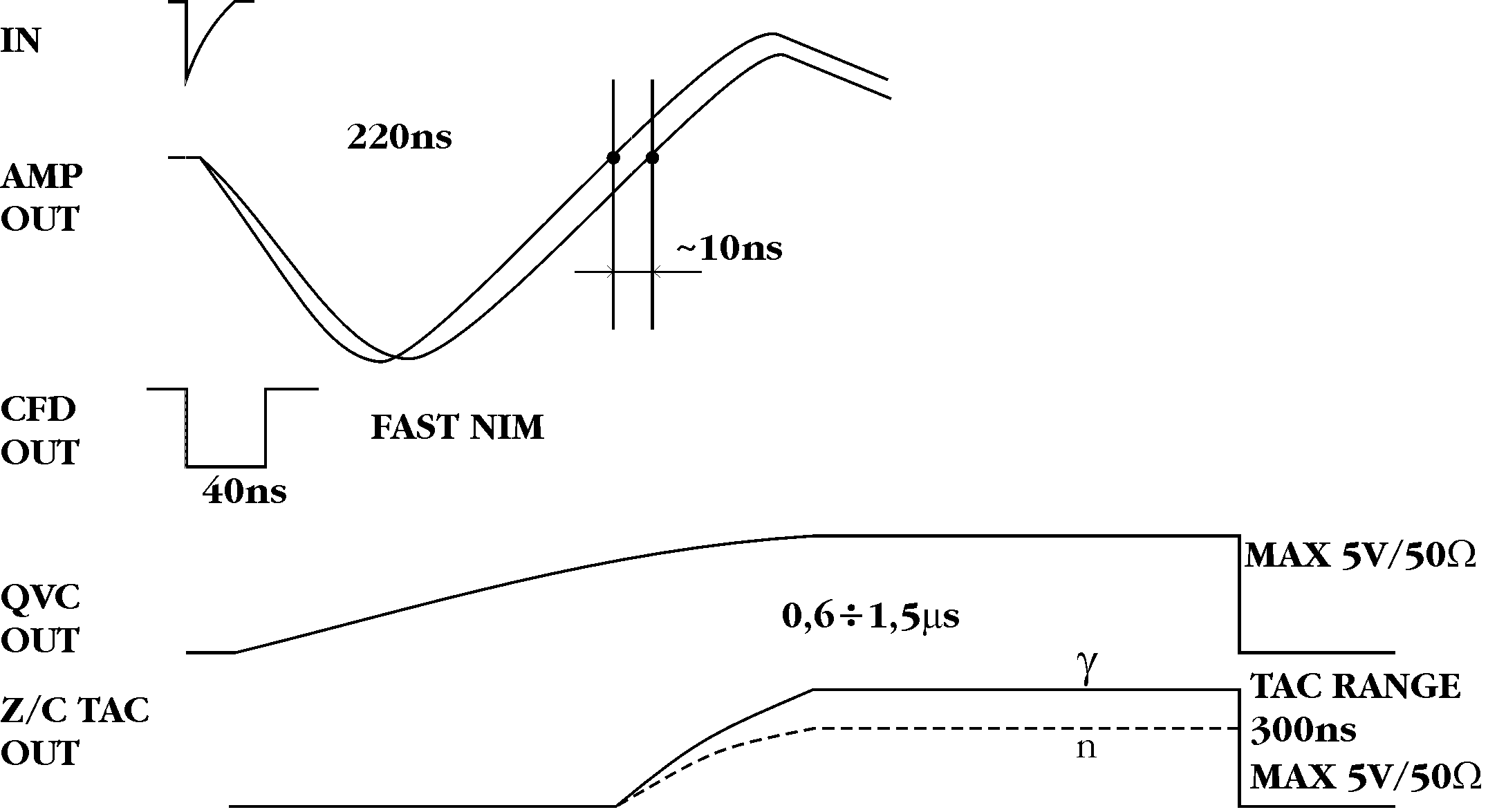}
 \caption[NDE202 timing diagram]{NDE202 timing diagram. Partial figure from \cite{NDE202_manual}.}
 \label{fig:przebieg}
\end{figure}

The \ac{PSD} in the Neutron Wall is done by NIM electronic units of the type NDE202 \cite{NDE202} based on the \ac{ZCO} discrimination technique. A block scheme of the NDE202 \ac{PSD} unit is shown in fig.~\ref{fig:nde202block}.
Several different outputs are available from the NDE202; for a complete description see ref.~\cite{NDE202_manual}. In this work the most interesting outputs are the \ac{QVC} and the \ac{ZCO} illustrated in fig.~\ref{fig:przebieg}. The \ac{TAC} signal between the trigger and the NDE202 \ac{CFD} is also an important output for \ac{TOF} measurements.
The signal from the \ac{PMT} anode is first split into a \ac{QVC} for energy measurements, a \ac{CFD} for timing and a shaping amplifier for the \ac{ZCO} discrimination, see section~\ref{ss:disc}. The \ac{QVC}, amplifier and \ac{CFD} signals are available at the front panel and the \ac{CFD} is used for the \ac{ZCO} discrimination. The output of the \ac{ZCO} block is both sent to an internal \ac{TAC} and to the front panel to optionally be used with an external \ac{TAC}.

\section{Status of the Neutron Wall\label{sec:status}}

\begin{figure}
 \centering
 \includegraphics[width=\textwidth]{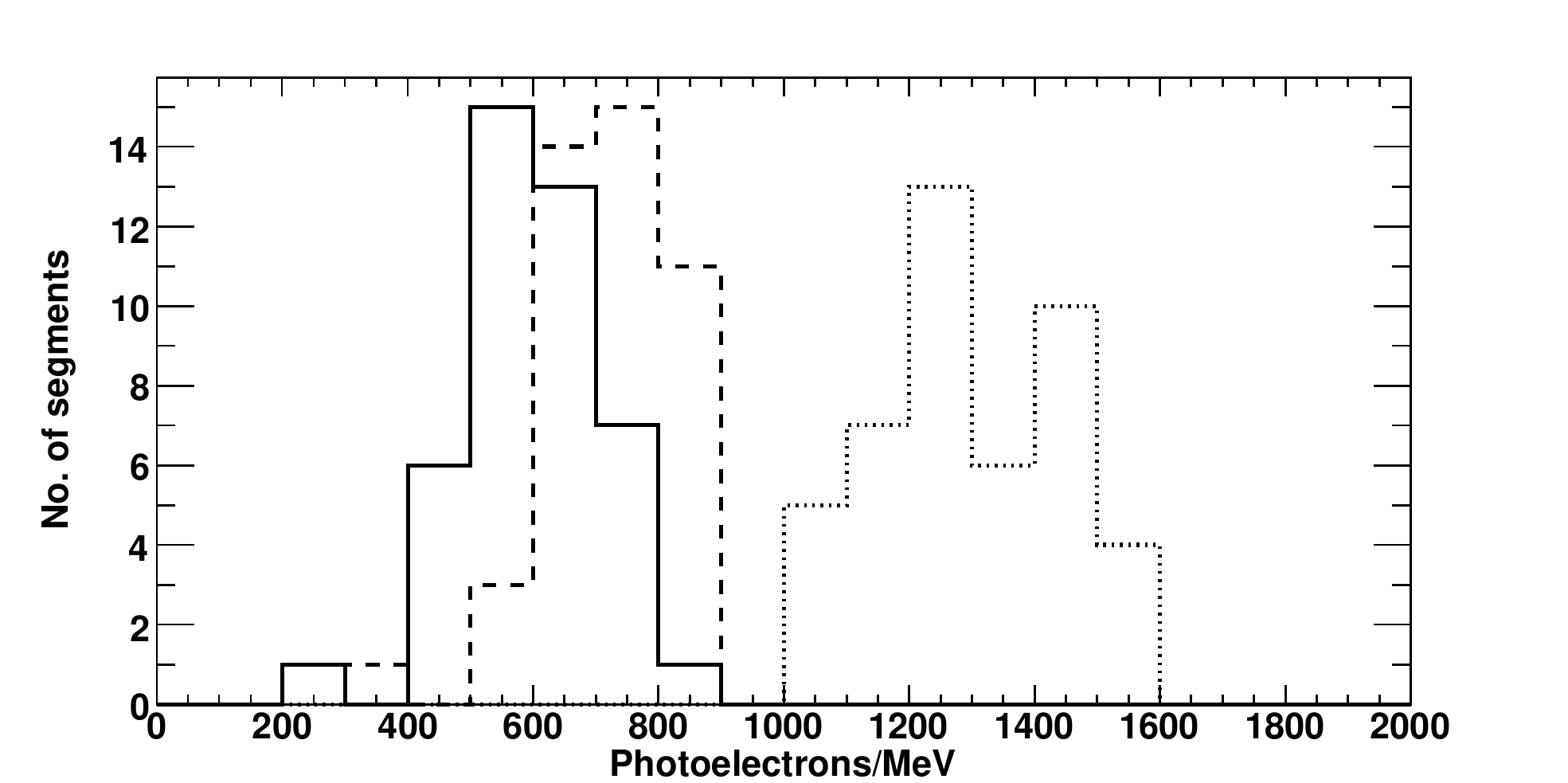}
 \caption[Distributions of photoelectrons per MeV from 1999, 2003 and 2009]{Distributions of photoelectrons per MeV from 1999 (dotted) as reported in \cite{1999NIMPA.421..531S}, 2003 after the repair (short dashed) as reported in \cite{howto_nphe} and 2009 (solid).}
 \label{fig:phemev}
\end{figure}

In preparation of future experiments with the Neutron Wall \cite{EBo,E482} the status of the Neutron Wall was checked in the beginning of 2009. In all measurements the same NDE202 unit (serial number 15) was used. Two important performance properties of the Neutron Wall is the detection efficiency and the quality of the separation between neutrons and $\gamma$ rays. A high efficiency is important in order to identify as many events as possible with one or several neutrons emitted. A good separation of neutrons and $\gamma$ rays is important in order to identify weak neutron-evaporation reaction channels. A property of the detector that is fundamental for both of these parameters is the number of photoelectrons that are generated per MeV of deposited energy. A high number of photoelectrons per MeV is important both to record low-energy neutron signals and to get good statistics in the slow component of the pulse for \ac{PSD} of neutrons and $\gamma$ rays. The number of photoelectrons per MeV was measured using the ratio between the Compton edge of $^{137}$Cs and the single photoelectron peak, as described in ref.~\cite{howto_nphe}. The result is shown in fig.~\ref{fig:phemev}.
As seen in fig.~\ref{fig:phemev}, the number of photoelectrons per MeV decreased substantially compared to 1999. The decrease occurred after a complete repair of the detectors in 2003. The reason for the large decrease of the number of photoelectrons per MeV after the repair is not understood. In 2005 the array was moved from IReS Strasbourg to GANIL. Segments 33 and 44 have shown a very low number of photoelectrons per MeV both in 2003 and in 2009. Assuming Gaussian distributions the average number of photoelectrons per MeV was $1280\pm50$ in 1999, $743\pm25$ in 2003 and  $597\pm21$ in 2009. The average number of photoelectrons per MeV has decreased by $42\pm5$\% and $20\pm4$\% between 1999 and 2003 and between 2003 and 2009, respectively.

\subsection{Efficiency Characteristics}

\begin{figure}
 \centering
 \includegraphics[width=0.9\textwidth]{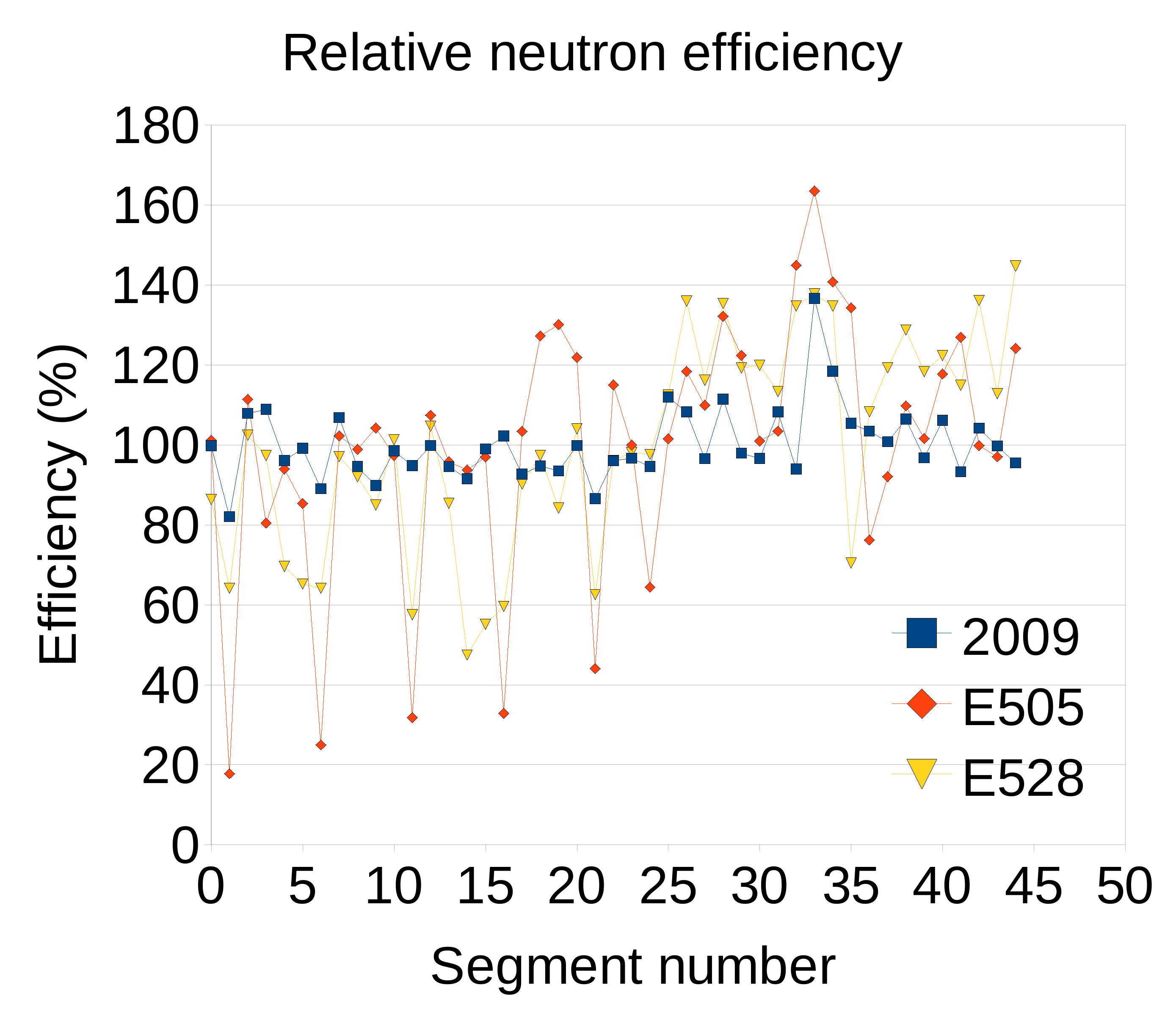}\\
 \includegraphics[width=0.9\textwidth]{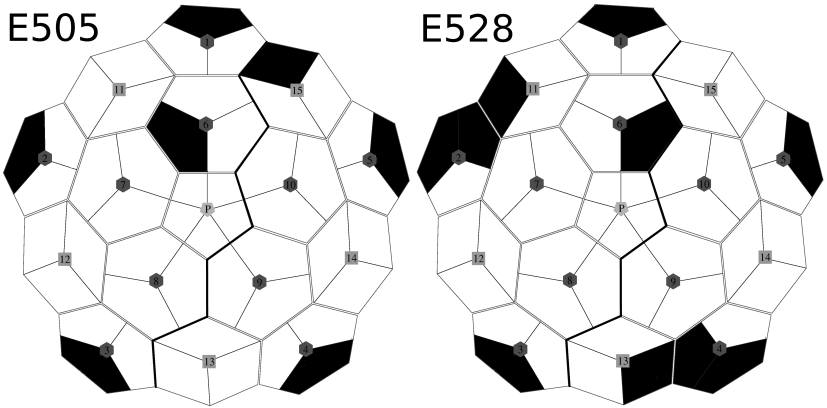}\\
 \caption[Relative efficiencies of the Neutron Wall segments]{Relative efficiencies of the Neutron Wall segments (top) from the $^{252}$Cf source tests in 2009, the EXOGAM + DIAMANT + Neutron Wall experiment E505 in 2006 and the SiCD + SiLi + EDEN + Neutron Wall experiment E528 in 2006. The E505 data is taken using the $^{36}$Ar + $^{40}$Ca/$^{16}$O fusion-evaporation reaction and the E528 data is taken using the $^{12}$C + $^{58}$Ni fusion-evaporation reaction. Drawings of the segments with a relative efficiency less than 80\% in black are also shown for the two experiments (bottom).
}
 \label{fig:relativeeffs}
\end{figure}

One of the problems with the Neutron Wall setup at GANIL is that the efficiencies of the outermost detector segments 1, 6, 11 and 16 sometimes have dropped radically. The reason for this is still unknown, but it is believed to origin from shadowing either by the supporting structures or by the beam pipe. In fig.~\ref{fig:relativeeffs} two examples of the shadowing effect are shown.
One of the examples is EXOGAM in its normal configuration with DIAMANT \cite{Scheurer1997501} and Neutron Wall as ancillary detectors (experiment E505). Using this setup the efficiency of the shadowed detectors has dropped to 20--40\%. The other example is experiment E528, using a setup with the Neutron Wall, EDEN \cite{eden}, and silicon charged particle detectors. The same detectors are still shadowed but this time the efficiency drops only to 50--70\%. The loss in efficiency due to the shadowing is about the same for neutrons and $\gamma$ rays which means that the material must be about equally good at stopping both of these particle species. To test whether this actually was an intrinsic problem with the Neutron Wall or a problem with shadowing from other surrounding material the relative efficiency for the segments were measured using only a $^{252}$Cf source and the Neutron Wall with no surrounding material. The results from this measurement is also shown in fig.~\ref{fig:relativeeffs} and the relative efficiency variance is only $\sim9$\% with no decrease in efficiency for the outermost segments. The conclusion is that the drop in efficiency is not an intrinsic problem with the Neutron Wall but that the problem is in the surrounding structures.

\subsection{Discrimination Between Neutrons and $\gamma$ rays}

\begin{figure}
 \centering
 \includegraphics[width=\textwidth]{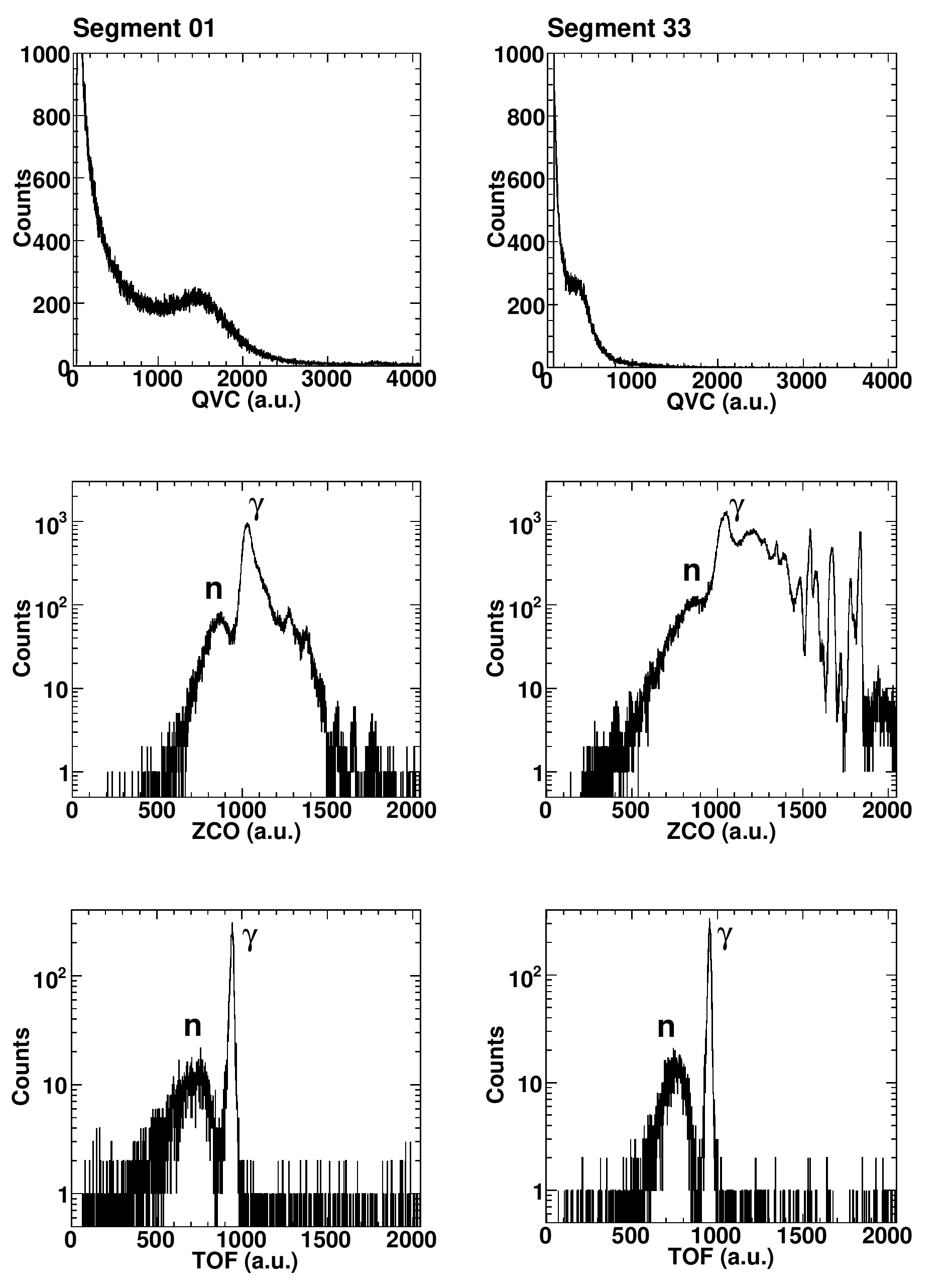}
 \caption[QVC spectrum, ZCO separation and TOF separation for two typical detector segments]{QVC spectrum measured by a $^{137}$Cs source (top), ZCO (middle) and TOF spectrum (bottom) measured by a $^{252}$Cf source for a typical detector segment which shows good performance (left) and one which shows bad performance (right). The time reference of the TOF measurement was obtained using a $2\times2$ inch BaF$_2$ detector \cite{paulas}. The sharp peaks on the right hand side of the ZCO spectrum for segment 33 are due to triggering on noise in the NDE202 unit.}
 \label{fig:twosegments}
\end{figure}

As expected, the quality of the separation between neutrons and $\gamma$ rays show the same trend as the number of photoelectrons per MeV for the detectors. All detector segments except number 33 and number 44 show a qualitatively similar separation both for \ac{TOF} and \ac{ZCO}. Quantitatively a \ac{FOM} defined as \cite{fom}
\begin{equation} \label{eq:FOMdef}
  \mathrm{FOM}=\frac{|X_{\gamma}-X_{\mathrm{n}}|}{W_\gamma +
    W_{\mathrm{n}}}.
\end{equation}
gives $\mathrm{FOM}\approx1$ for the good segments and $\mathrm{FOM}\approx0.5$ for the bad segments \cite{paulas}.
This \ac{FOM} is a unit-less ratio of the difference between the peak positions $X_i$ divided by the sum of their \ac{FWHM}, $W_i$ ($i=\gamma,n$). An increased value of the \ac{FOM} corresponds to a better discrimination between neutrons and $\gamma$ rays. In fig.~\ref{fig:twosegments} the separation between neutrons and $\gamma$ rays are shown for two detector segments: number 1 and number 33.

\subsection{Effects of Varying the PMT High Voltages}

\begin{figure}
 \centering
 \includegraphics[width=\textwidth]{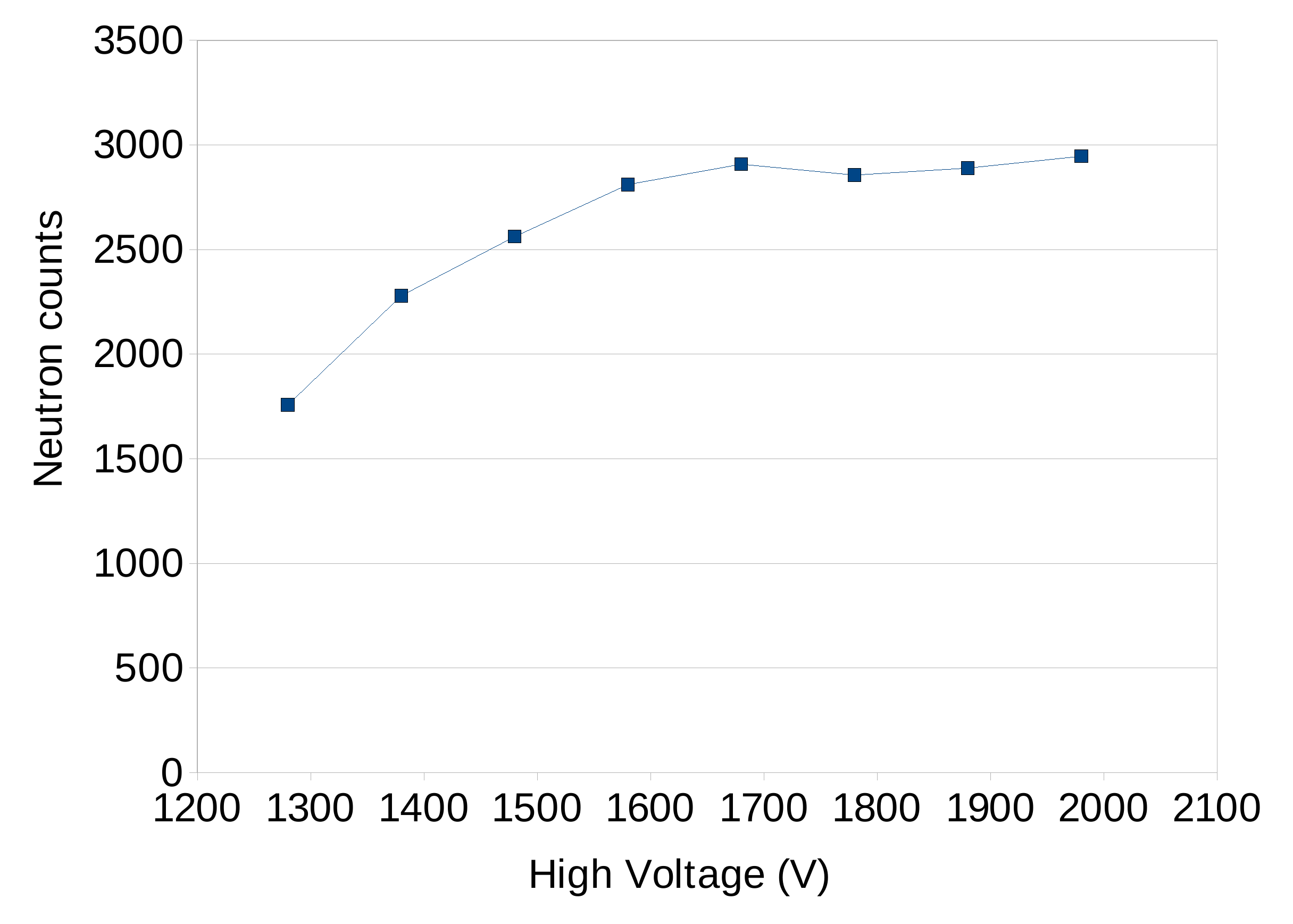}
 \caption[Number of counts in the neutron peak as a function of the high voltage for detector segment number 1]{Number of counts in the neutron peak as a function of the high voltage for detector segment number 1. The increase in relative efficiency is about 60\%.}
 \label{fig:HV}
\end{figure}

The \ac{HV} values of the Neutron Wall \ac{PMT}s are set to values below their design values. The \ac{PMT}s in the hexagonal detectors are designed for a maximum \ac{HV} of $-2700$~V, the \ac{PMT}s in the pentagonal detectors for a \ac{HV} of $-3300$~V while in the Neutron Wall the \ac{HV} values are set to an average of $-1390$~V and $-2160$~V, respectively. The reason for this is a problem with the dynamic range of the \ac{ZCO} system of the NDE202 units. The quality of the \ac{ZCO} \ac{PSD} becomes worse when higher voltage is used. Many interactions in the detectors, however, lead to low kinetic energy transfer. A detection of more of such low-energy neutron interactions would lead to a gain in efficiency. Therefore, one would preferably like to use a higher voltage for the \ac{PMT}s as shown in fig.~\ref{fig:HV}.

\begin{figure}
 \centering
 \includegraphics[width=\textwidth]{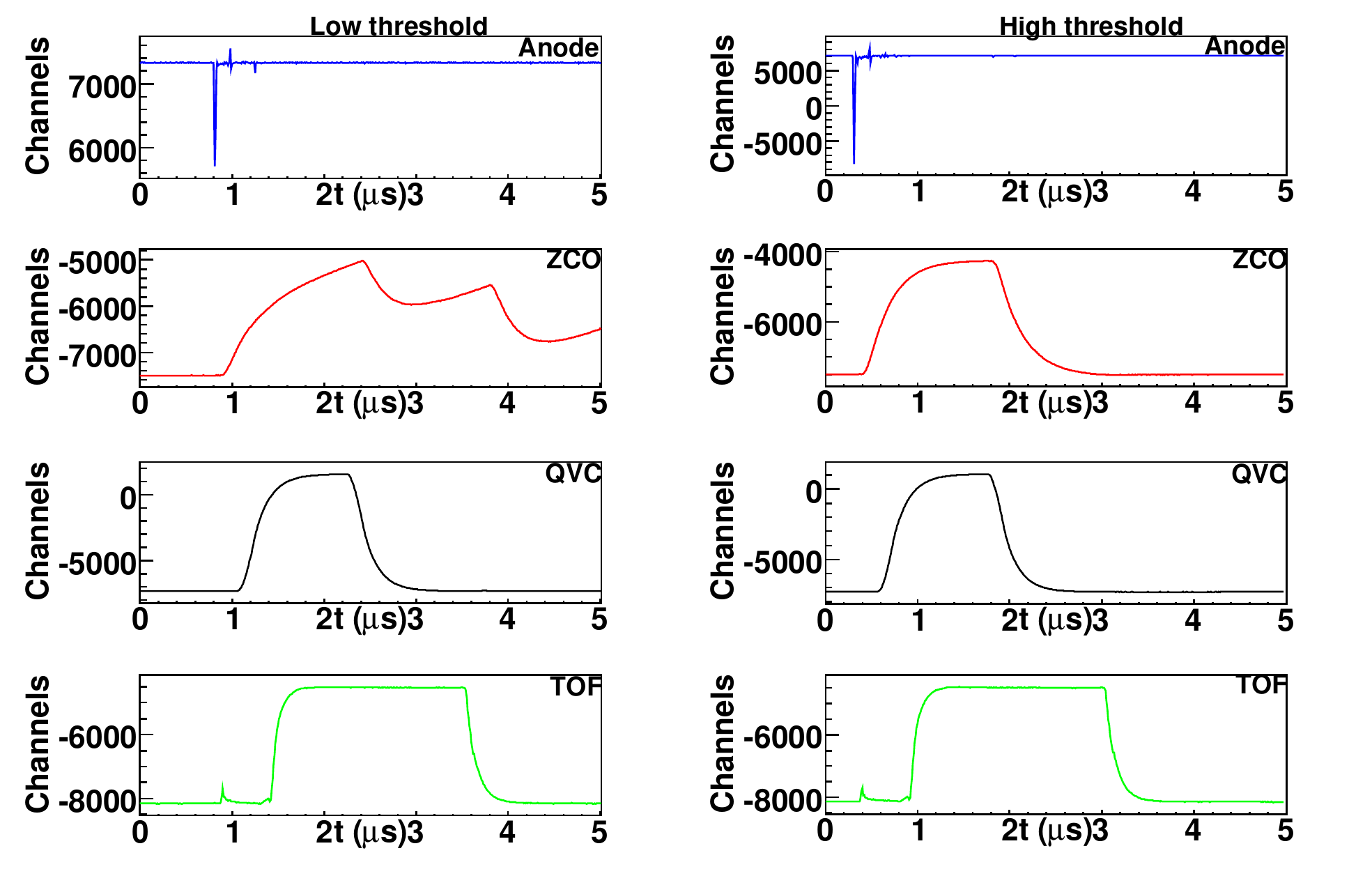}
 \includegraphics[width=\textwidth]{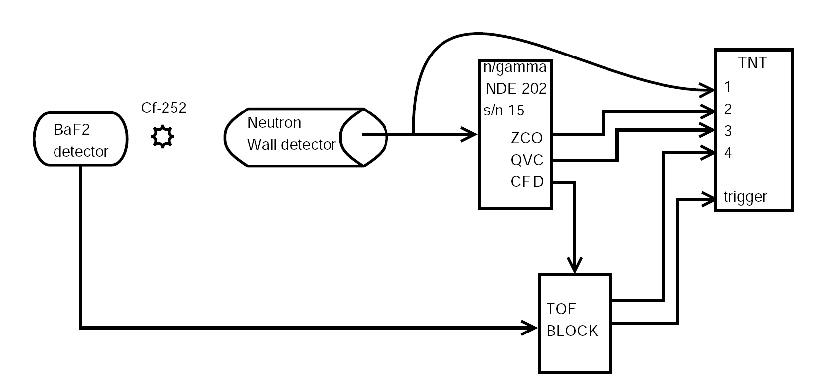}
 \caption[QVC, ZCO, TOF and anode signals from detector segment 1 with low threshold and high threshold]{QVC, ZCO, TOF and anode signals from detector segment 1 with low threshold (left) and high threshold (right). This data was taken by using a TNT2 module. The lower part of the figure, showing the setup of the measurement, is from ref.~\cite{paulas}.}
 \label{fig:pulseshapes}
\end{figure}

To keep as many low-energy neutron interactions as possible and to compensate for the lower \ac{HV} and thus smaller signals one has to set the threshold of the \ac{CFD} lower. There is however a problem with using too low threshold in the NDE202, since this causes the signals from the \ac{QVC} and \ac{ZCO} to oscillate in a strange way as shown in fig.~\ref{fig:pulseshapes}.

\begin{figure}
 \centering
 \includegraphics[width=\textwidth]{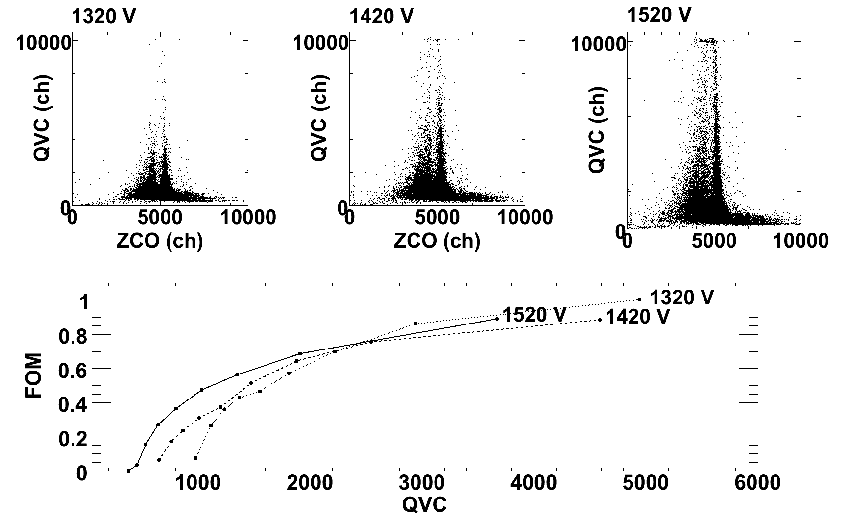}
 \caption[ZCO separation of neutrons and $\gamma$ rays versus QVC and FOM values as a function of QVC for different HV settings]{ZCO separation of neutrons and $\gamma$ rays versus QVC (top) and FOM values as a function of QVC (bottom) for different HV settings. The QVC values of the FOM for the 1320~V and 1420~V data were calibrated to correspond to the same deposited energy as in the 1520~V data.}
 \label{fig:hvs}
\end{figure}

Thus, there is two main objectives for increasing the \ac{HV} of the \ac{PMT}: to increase the efficiency and to be able to use \ac{CFD} thresholds that have some margin to the lower limit of operation.
To study the effect of an increasing \ac{HV} on the \ac{PSD} quality, data sets were taken for segment 33 using three different \ac{HV} settings: 1320~V, 1420~V and 1520~V. The results of these measurements are shown in fig.~\ref{fig:hvs}.
The \ac{ZCO}, \ac{QVC} and \ac{TOF} data was stored on an event-by-event basis and the \ac{FOM} was calculated for each \ac{HV} setting and for ten different ranges of the \ac{QVC} that were selected to contain 1/10 of the events in each the data set. When comparing the \ac{FOM}, the \ac{QVC}s were calibrated relative to each other using two measurements of the $^{137}$Cs Compton edge with \ac{HV} settings of 1520~V and 1670~V. Assuming an exponential gain, a fit of the form
\begin{equation}
 y = a2^{\frac{x}{b}}
\end{equation} 
was obtained with $y$ being the \ac{QVC} in channel numbers, $a$ the normalization, $x$ corresponding to the voltage and $b = 158$ obtained from the fitted value of the \ac{HV} to double the gain from the \ac{PMT}. The three \ac{FOM} curves in fig.~\ref{fig:hvs} look roughly similar and one can conclude that the decrease in separation quality between neutrons and $\gamma$ rays in the one-dimensional projections mainly seem to be due to more low-energy neutron interactions. The \ac{FOM} is even better at fixed calibrated \ac{QVC} for higher \ac{HV}. Thus one should be able to increase the efficiency of the Neutron Wall by increasing the \ac{HV} of the \ac{PMT}s if the \ac{ZCO} is used together with the \ac{QVC} and \ac{TOF} in the discrimination procedure between neutrons and $\gamma$ rays.
\cleardoublepage

\chapter{Pulse-Shape Analysis with Digital Electronics\label{sec:dpsd}}
For the development of the next generation of neutron detector arrays one should take advantage of digital electronics. Until now, the \ac{PSD} of neutrons and $\gamma$ rays has been made by utilizing analogue electronic methods \cite{alexander1961,roush1964psd,brooks}. With the fast development of digital electronics one is no longer limited to using only analogue components. The term digital electronics means, in this context, that the detector signal is digitized with a fast sampling \ac{ADC} and then processed using a programmable device  like a \ac{DSP}, \ac{FPGA} or even a \ac{PC}. This means that one can use much more sophisticated \ac{PSA} algorithms than with analogue electronics. But digital \ac{PSA} also has limitations regarding, for example, computing time and signal reconstruction. This chapter is based on ref.~\cite{dpsd} and concerns the influence of the bit resolution and sampling frequency of the \ac{ADC} on the \ac{PSD}. The experiment and results of the \ac{PSD} of neutrons and $\gamma$ rays is described in section~\ref{ss:dspmeth} and the effects of the sampling frequency on the time resolution is presented in section~\ref{ss:timeres}.

\section{Digital PSD Methods\label{ss:dspmeth}}

Several sophisticated methods for digital \ac{PSD} have been developed by various research groups lately. One of these methods uses a previously measured standard pulse shape from the detector. By defining a correlation function of a sample pulse shape and the standard pulse shape one can discriminate between neutrons and $\gamma$ rays \cite{2003NIMPA.497..467K}. Another method that uses standard pulse shapes has also been developed and is being implemented in a \ac{FPGA} \cite{2002NIMPA.490..299T,Guerrero2008212}. In this method the measured pulse shape is fitted with ``true'' neutron and $\gamma$-ray pulse-shapes and the $\chi^2$ values of these fits are compared. A third method that has been developed for digital electronics is the pulse gradient method \cite{2007NIMPA.578..191D,2007NIMPA.583..432A}. In this method two sampling points from the tail of the pulse are selected and the slope between these two points is calculated. All these methods have yielded good results regarding the discrimination of neutrons and $\gamma$ rays. 

In ref.~\cite{dpsd} two methods, as described below, have been developed to perform digital \ac{PSD} and to study the effects of the \ac{ADC} bit resolution and sampling frequency on the \ac{PSD} quality. These methods were developed to be numerically simple, such that the limitations from computing time should be not too large. They were also selected due to their similarity with well studied analogue methods.

\paragraph{Charge Comparison}

\begin{figure}
 \begin{center}
 \includegraphics[width=\columnwidth]{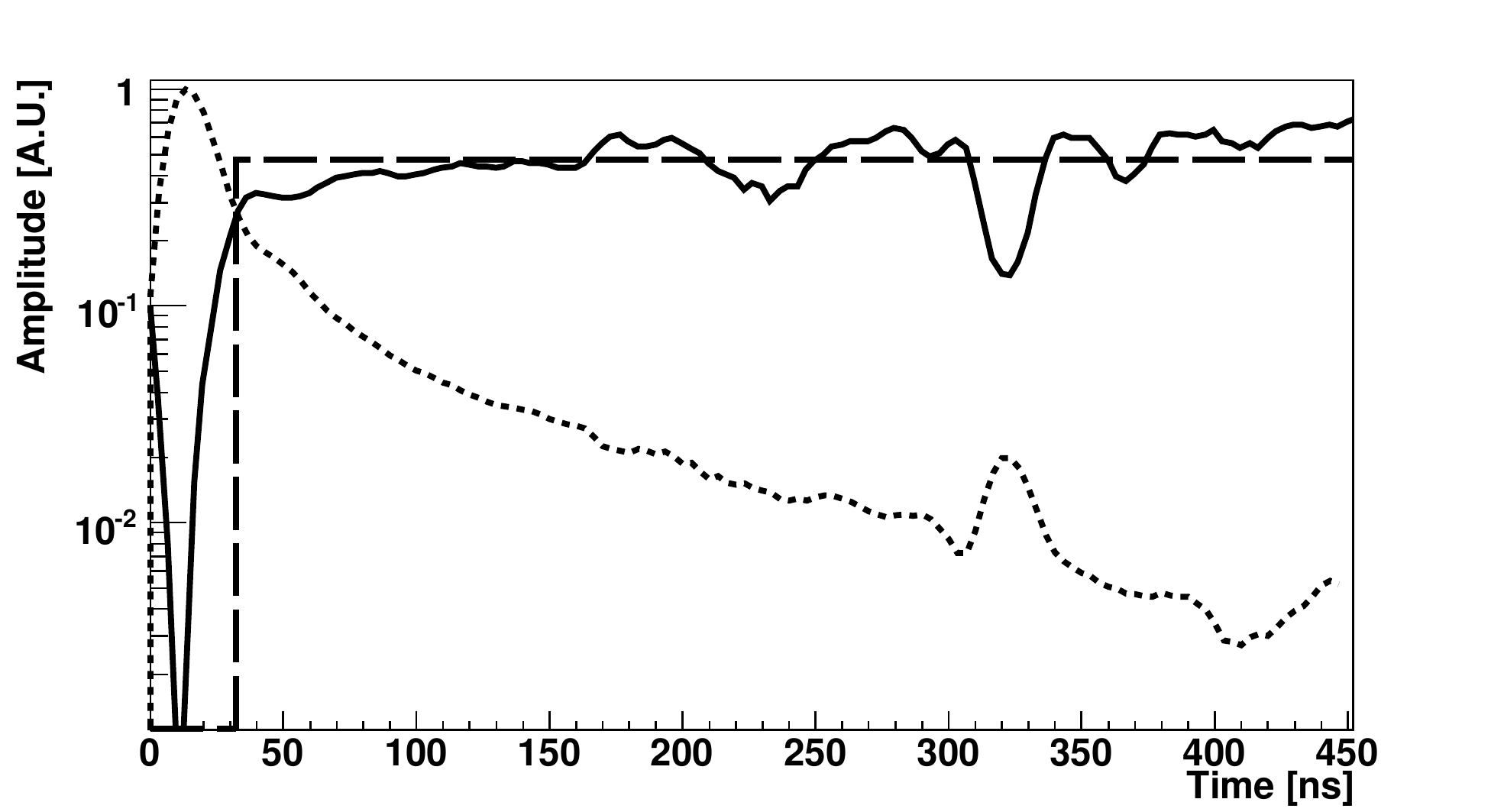}
\end{center}
\caption[Weighting function $w(t)$]{Weighting function $w(t)$ for digital (solid) and analogue (dashed) charge comparison PSD shown together with an average neutron pulse (short-dashed). Reprinted from \cite{dpsd} with permission from Elsevier.\label{fig:wt}}
\end{figure}

In the charge comparison method two integration gates are set on the fast and slow decay components of the pulse. By comparing these integrals with each other one will get a separation between neutrons and $\gamma$ rays. Using digital electronics this method can be generalized to concern evaluating the integral
\begin{equation} \label{eq:psdobs}
 S = \int_{0}^{T} p(t) w(t) \sd t,
\end{equation}
where $T$ is the time to evaluate the pulse $p(t)$, $w(t)$ is a weighting function and $S$ is a quantity that differs between neutrons and $\gamma$ rays. To enhance the \ac{PSD} maximally, the best choice of $w(t)$ has been shown \cite{gattimartini} to be 
\begin{equation} \label{eq:optimalw}
  w(t) =
  \frac{\bar{n}(t)-\bar{\gamma}(t)}{\bar{n}(t)+\bar{\gamma}(t)},
\end{equation}
where $\bar{n}(t)$ and $\bar{\gamma}(t)$ are the average neutron and $\gamma$-ray pulse shapes, respectively. See fig.~\ref{fig:wt} for an example of how $w(t)$ looks like for the experiment in ref.~\cite{dpsd}.

\paragraph{Zero Cross-over}

\begin{figure}
 \centering
 \includegraphics[width=\textwidth]{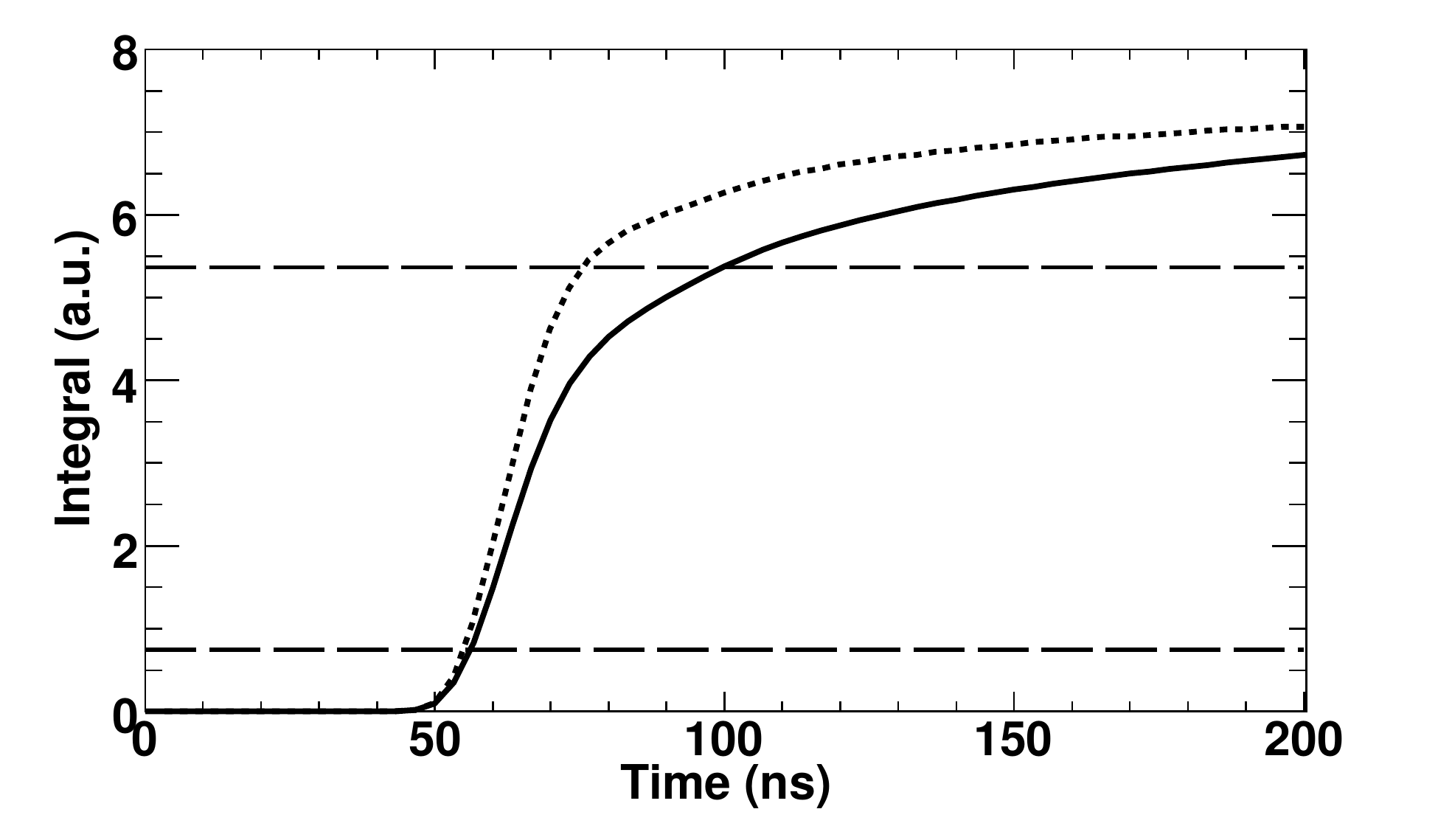}
 \caption[Difference between the integrated rise
    time of a $\gamma$-ray and a neutron pulse]{Difference between the integrated rise
    time of a $\gamma$-ray (dashed) and a neutron (solid) pulse. 
    The points at 10 \% and 72 \% of the pulse height are 
    indicated by long-dashed lines.}
 \label{fig:rise}
\end{figure}

The other analogue method that has been adapted digitally is the \ac{ZCO} method. The \ac{ZCO} method is described in sections~\ref{ss:disc} and~\ref{ss:psd}. Integrating the pulse and taking the rise time of the integrated pulse has been shown to be equivalent to shaping and taking the zero-crossing time \cite{nima354_380}. In fig.~\ref{fig:rise} the principle of the integrated rise-time method is shown. It was found that for the setup in ref.~\cite{dpsd} the best result was obtained by using the 10--72\% rise time.

\subsection{Experimental Setup\label{ss:expset}}

\begin{figure}
 \centering
 \includegraphics[width=\textwidth]{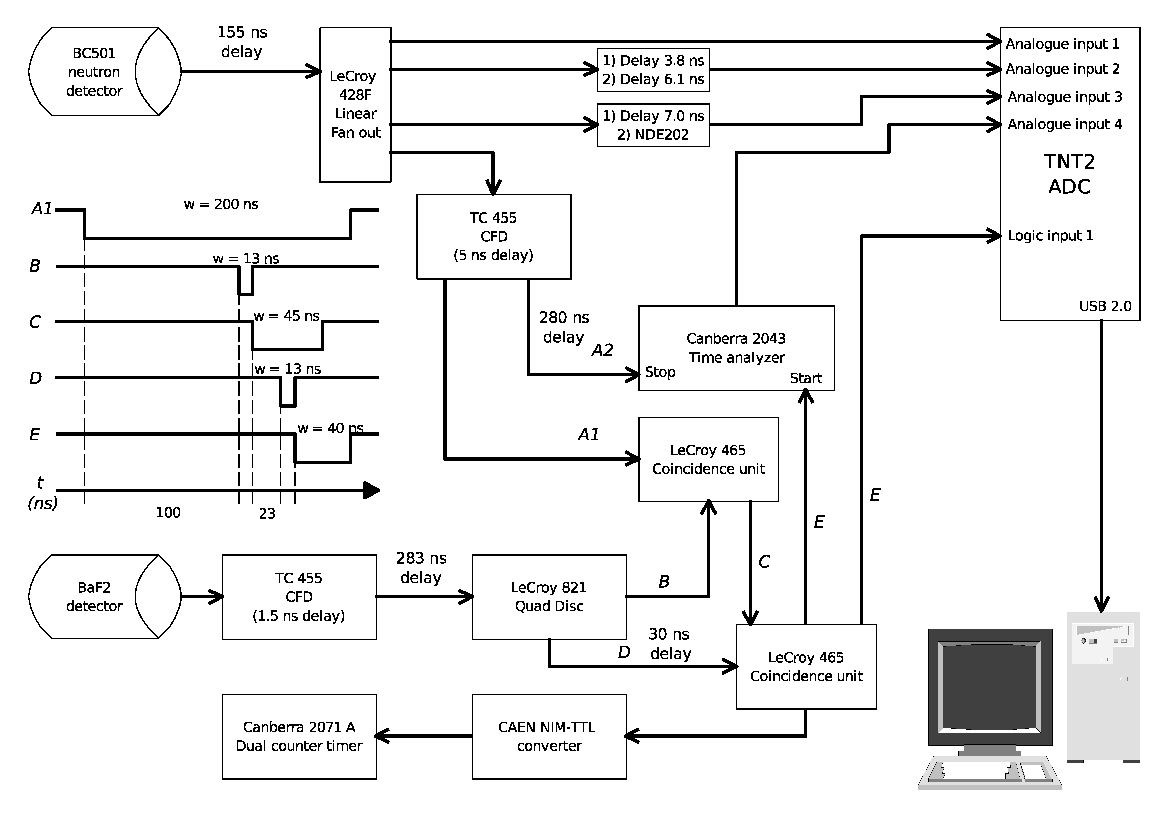}
 \caption[Schematic electronics block and timing diagrams of the setup]{Schematic electronics block and timing diagrams of the setup
    used in this work. The delay units consist of coaxial cables of type
    RG58 (delays $>10$ ns) or RG174. The signals shown in the timing
    diagram correspond to the detection of prompt $\gamma$ rays in the
    two detectors.
    1) 300 MS/s and
    2) 200 MS/s setup. Reprinted from \cite{dpsd} with permission from Elsevier.}
 \label{fig:setup}
\end{figure}

To test the digital \ac{PSD} methods, an experiment was carried out using a $^{252}$Cf source, a liquid scintillator detector
 from the NORDBALL neutron detector array \cite{1991NIMPA.300..303A}, a BaF$_2$ detector and a TNT2 sampling \ac{ADC} system running at 300~\ac{MS/s} and 14~bits \cite{tnt_ieee}. See fig.~\ref{fig:setup} for a drawing of the setup and ref.~\cite{dpsd} for a detailed description.

\begin{figure}
 \centering
 \includegraphics[width=\textwidth]{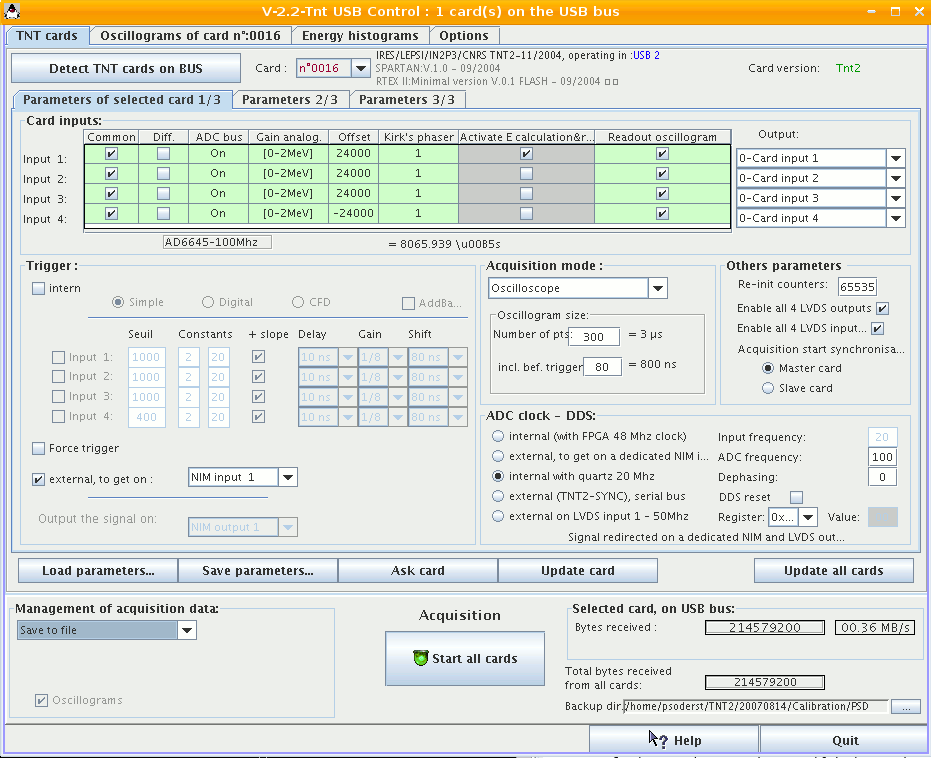}
 \caption[The TNT USB Control graphical user interface]{The TNT USB Control graphical user interface and the settings used in the 300~\ac{MS/s} experiment. }
 \label{fig:tuc}
\end{figure}

\begin{figure}
 \centering
 \includegraphics[width=\textwidth]{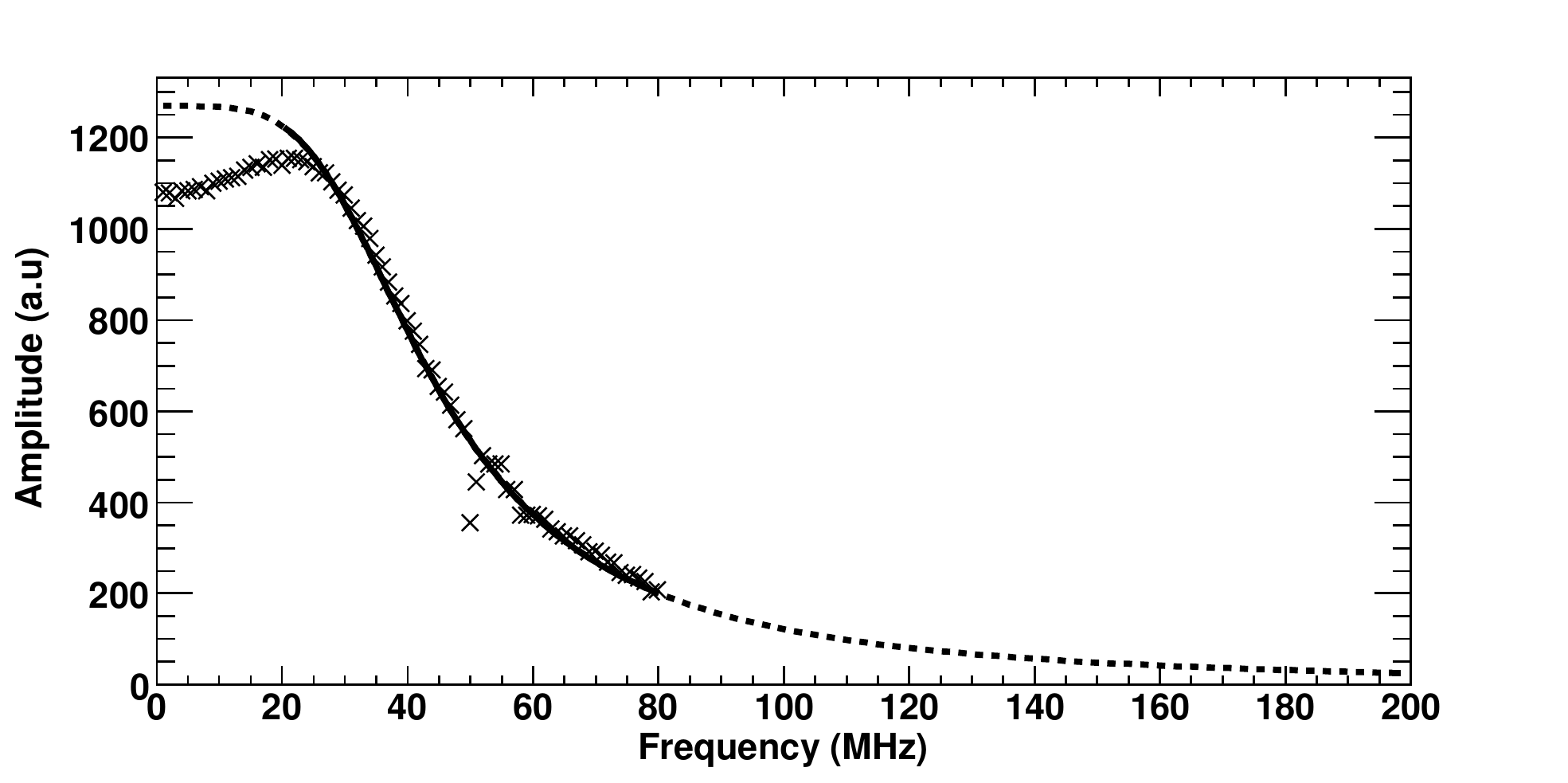}
 \caption[Frequency response of the system]{Frequency response of the system measured up to 80~MHz (crosses) together with the fitted (solid) and extrapolated (dashed) low-pass filter functions.}
 \label{fig:freqresponse}
\end{figure}

\begin{figure} [ht]
  \centering
 \includegraphics[width=\columnwidth]{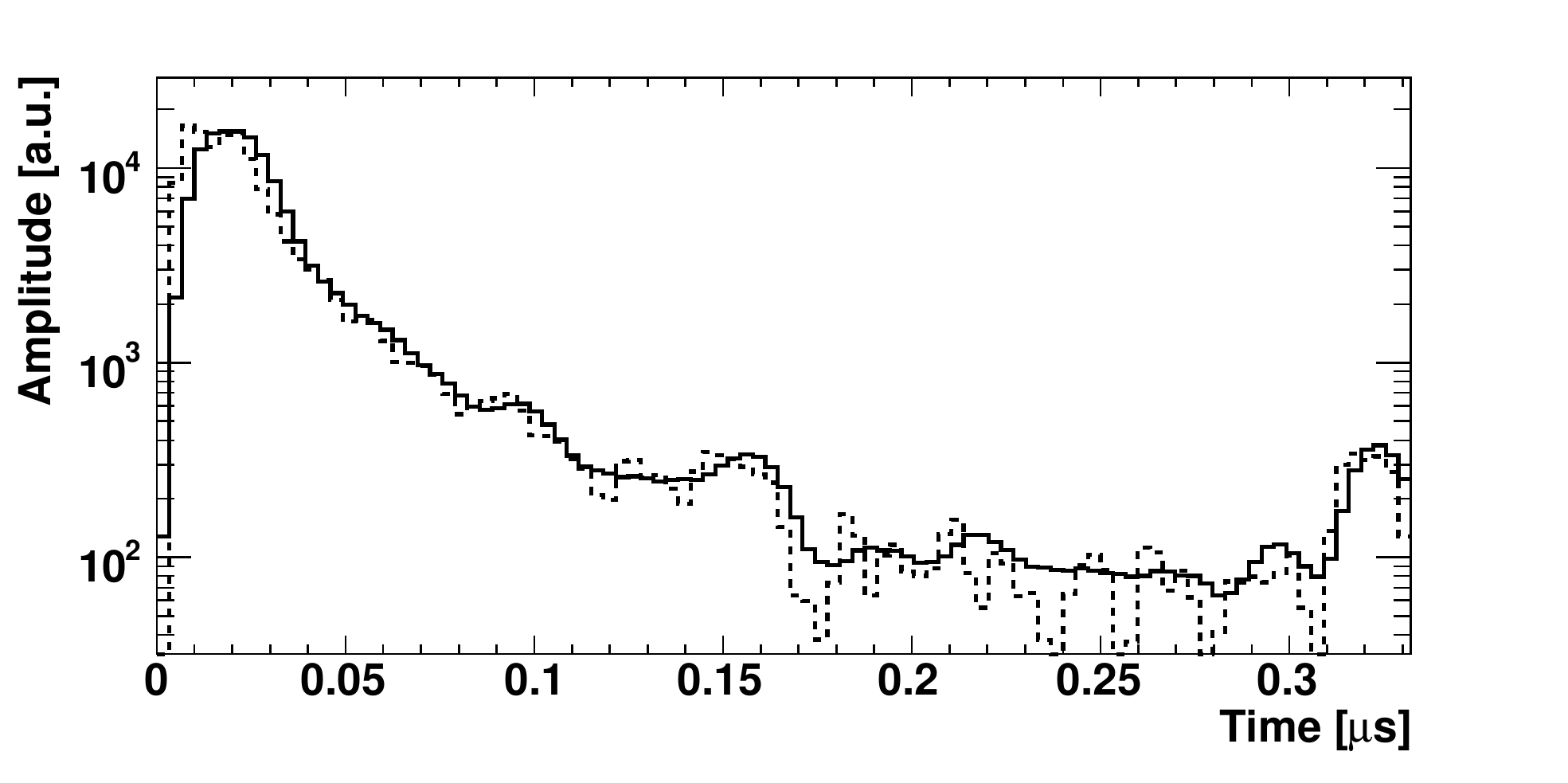}
  \caption[Effects of the limited analogue bandwidth]{Effects of the limited analogue bandwidth. A typical
    measured pulse shape (solid histogram) and the corresponding
    undistorted pulse shape after correcting for the finite analogue
    bandwidth (dashed histogram) are shown. To remove
    artifacts due to the finite size of the discrete Fourier
    transform, the histograms are smoothed twice using the 353QH
    algorithm \cite{353QH}. Reprinted from \cite{dpsd} with permission from Elsevier.\label{fig:analogue}}
\end{figure}

The TNT2 is a single width NIM unit with four independent channels. Each channel has a 14~bit and 100~\ac{MS/s} flash ADC and an analogue
bandwidth of 40~MHz. The TNT2 is set up, controlled and read out through a Java graphical user interface by a Linux \ac{PC} via a USB 2.0 interface, see fig.~\ref{fig:tuc}.
To increase the sampling frequency of the readout, three channels of the TNT2 were connected in parallel. With this technique it was possible to increase the sampling frequency from 100~\ac{MS/s} to 300~\ac{MS/s}. This technique does not, however, increase the analogue bandwidth of the setup. To study the effects of this the frequency response of the system was measured, see fig.~\ref{fig:freqresponse}.
The frequency response measurements was carried out by using an Agilent function generator of model
33250A. A sine function of a fixed amplitude was used as an input, $A_{\mathrm{in}}$, and the output amplitude, $A_{\mathrm{out}}$, was measured in the range 1--80~MHz. Above 80~MHz the frequency response was extrapolated using a fit of a general low pass filter function
\begin{equation}
 A_{\mathrm{out}}(f) = \frac{a_0}{\sqrt{1+\left(\frac{f}{f_{0}}\right)^{a_1}}},
\end{equation}
with $f$ the frequency, $a_0$ a free scaling parameter, $a_1=2.27$ the fitted value of the filter order and $f_{0}=35.7$~MHz the fitted cut-off frequency.
The effect of the analogue bandwidth was examined by doing a discrete Fourier transform, correcting for the frequency response and transforming back to the time domain. The result of this test is shown in fig.~\ref{fig:analogue} and the effect of the limited analogue bandwidth was judged to be small enough in order not to to contribute significantly to the final results.

Data sets were collected for two different setups, one setup with 300~\ac{MS/s} and one setup with 200~\ac{MS/s} and an analogue \ac{PSD} unit of the type NDE202, described in section~\ref{ss:psd}. To evaluate the performance of the \ac{PSD} methods for different properties of the \ac{ADC} the 300~\ac{MS/s} data set was reduced regarding bit resolution and sampling frequency to several other data sets. For the bit resolution an integer division by $2^{14-b}$ of the value of each sampling point value, where $b$ is the
number of bits in the new data set, was used. For the sampling frequency a number of data points
were removed, resulting in new data sets corresponding to sampling frequencies of 150~\ac{MS/s}, 100~\ac{MS/s}, 75~\ac{MS/s}, 60~\ac{MS/s} and 50~\ac{MS/s}.

\subsection{Figure-of-Merit}

\begin{figure}
 \centering
 \includegraphics[width=\textwidth]{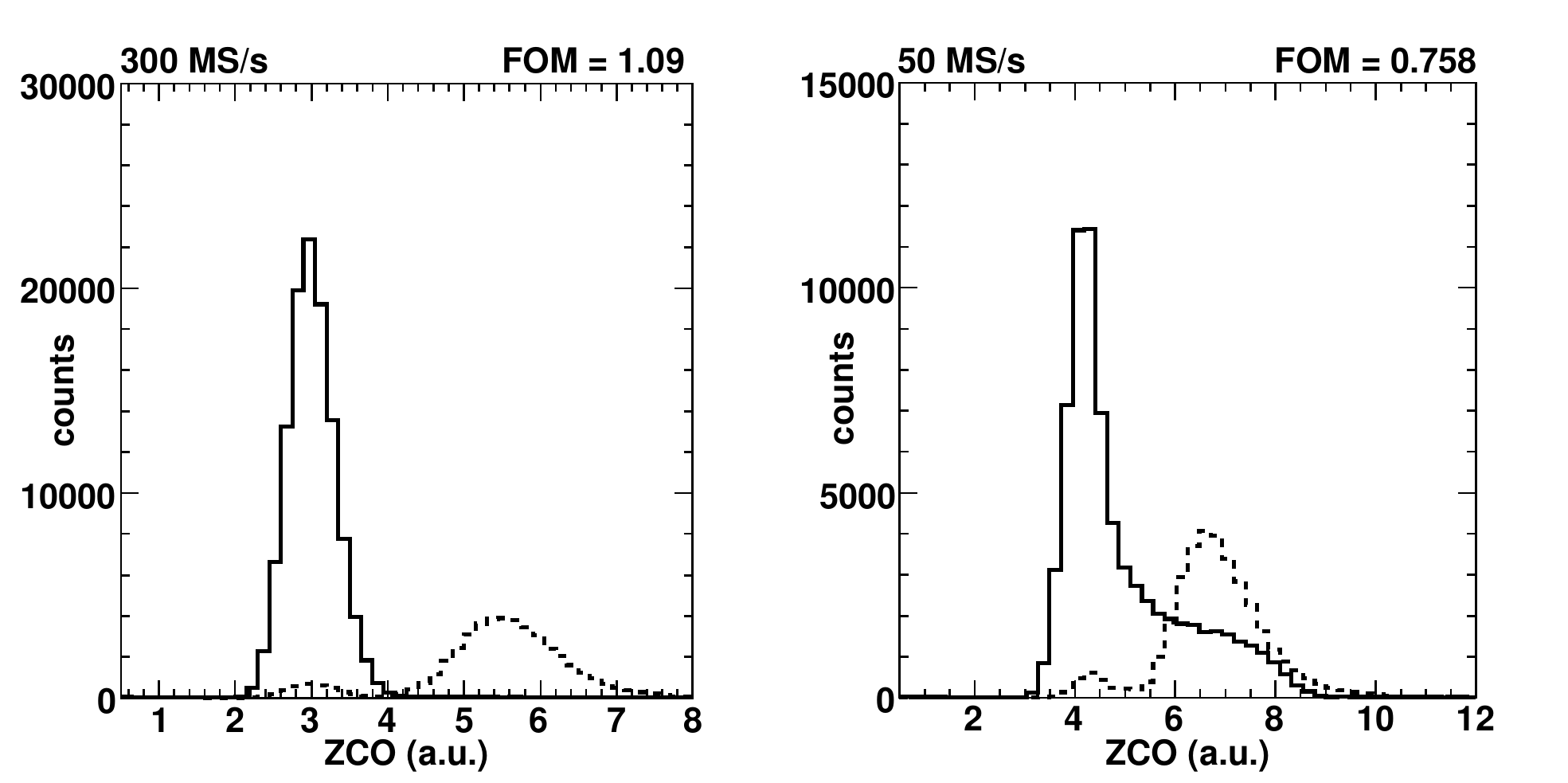}
 \caption[TOF gated neutron and $\gamma$-ray distributions for the ZCO digital PSD based method]{TOF gated neutron (dashed) and $\gamma$-ray (solid) distributions for the ZCO based digital PSD method and sampling frequencies 300~MS/s and 50~MS/s.}
 \label{fig:fomproblems}
\end{figure}

To quantify the results the \ac{FOM} in eq.~\ref{eq:FOMdef} was used. However, this \ac{FOM} has a few drawbacks. One drawback is that it only takes the peak positions and widths in account but not the background under the distributions. As seen in fig.~\ref{fig:ng2d} there are events originating from pile-up of
several closely spaced events and random $\gamma$ rays that, when projected to a one-dimensional distribution, will make up a continuous background. The \ac{FOM} in eq.~\ref{eq:FOMdef} also assumes Gaussian, or at least symmetric, distributions. If the distributions are not Gaussian or symmetric the \ac{FOM} value can be very misleading as shown in fig.~\ref{fig:fomproblems}.
The third drawback of \ac{FOM} in eq.~\ref{eq:FOMdef} is that it only works for one-dimensional distributions while, as mentioned in section~\ref{ss:disc}, in real experiments two-dimensional selection criteria are mostly used. To complement the \ac{FOM} a parameter, $R$, was defined as
\begin{equation} \label{eq:p_g}
R = \frac{N_\mathrm{b}}{N_\mathrm{n}-N_\mathrm{b}},
\end{equation}
based on the estimated number of $\gamma$-ray background counts, $N_\mathrm{b}$, relative to the number of neutron counts, $N_\mathrm{n}$, in the neutron peak. See ref.~\cite{dpsd} for details on how to calculate $R$.

\subsection{Results}

\begin{figure}
 \centering
 \includegraphics[width=\textwidth]{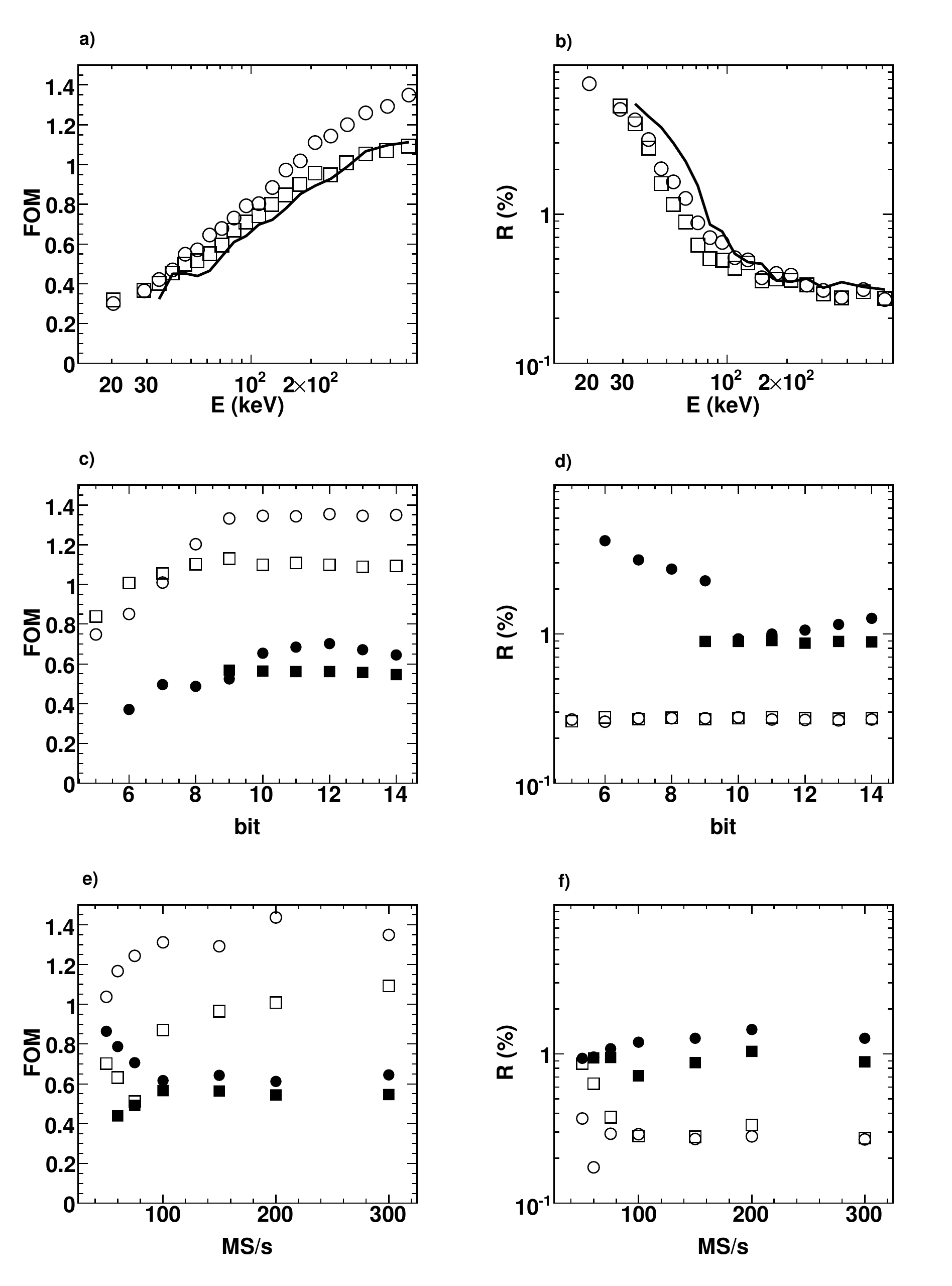}
\caption[FOM and $R$ for the ZCO and charge comparison based methods as a function of energy, bit resolution and sampling frequency]{FOM (a, c, e) and $R$ (b, d, f) for the ZCO (circles) and charge comparison (squares) based methods as a function of energy (a, b), bit resolution (c, d) and sampling frequency (e, f). The solid line (a, b) is from analogue reference data using NDE202 electronics. Filled and empty symbols (c, d, e, f) correspond to an electron (proton) energy gate of $E_\mathrm{e}=500$--$700$~keV ($E_\mathrm{p}=2200$--$2700$~keV) and $E_\mathrm{e}=50$--$57$~keV ($E_\mathrm{p}=500$--$540$~keV) respectively.\label{fig:fombitfreq}}
\end{figure}

The results from the digital \ac{PSD} are shown in fig.~\ref{fig:fombitfreq}.
As can be seen, the digital \ac{PSD} gives at least as good separation as the analogue \ac{PSD} in the entire energy range. The \ac{ZCO} based \ac{PSD} method is shown to work better than the charge comparison based. The \ac{FOM} saturates around 9 bits for the charge comparison based method and
at about 10 bits for the \ac{ZCO} based method. It should be noted that these values are for the dynamic range of this experiment with $\gamma$-ray energies between $E_\mathrm{e}=15$--$700$~keV and recoil proton energies between $E_\mathrm{p}=250$--$2700$~keV. Increasing the bit resolution by one unit would double the dynamic range which implies that a bit resolution of 12 bits would be adequate for most experiments since this allows for \ac{PSD} up to recoil proton energies of $E_\mathrm{p}=12$~MeV.
For the low energy pulses the \ac{FOM} shows no strong dependence of the sampling frequency above 100~\ac{MS/s}. For high energy pulses the \ac{FOM} increases slowly above 100~\ac{MS/s}. The $R$ values saturate already around 75~\ac{MS/s}. The apparently odd frequency behavior for low-energy pulses is due to the asymmetries shown in fig.~\ref{fig:fomproblems}.

\section{Time Resolution\label{ss:timeres}}

In a fully digital system the \ac{TOF} measurement should also be determined by the digitized pulse in the neutron detector. In ref.~\cite{dpsd} a test of the influence of the finite sampling frequency on the achievable time resolution was made. Two TNT2 channels, each recorded with a sampling frequency of 100 \ac{MS/s}, were compared and a timing parameter $\Delta t_{21}$ was defined as the difference between the extracted leading edge times of each of the two channels. The distribution of $\Delta t_{21}$ was used as an estimate of the time resolution due to the finite sampling frequency. A \ac{FWHM} of 1.7 ns was extracted and by comparing this to the achievable intrinsic time resolution of a liquid scintillator detector plus \ac{PMT}, typically \ac{FWHM}~=~1.5~ns or larger, it was concluded that the contribution of the finite sampling frequency to the total FWHM of the time resolution should be almost negligible already at 200~\ac{MS/s}.

\begin{figure}
 \centering
 \includegraphics[width=\textwidth]{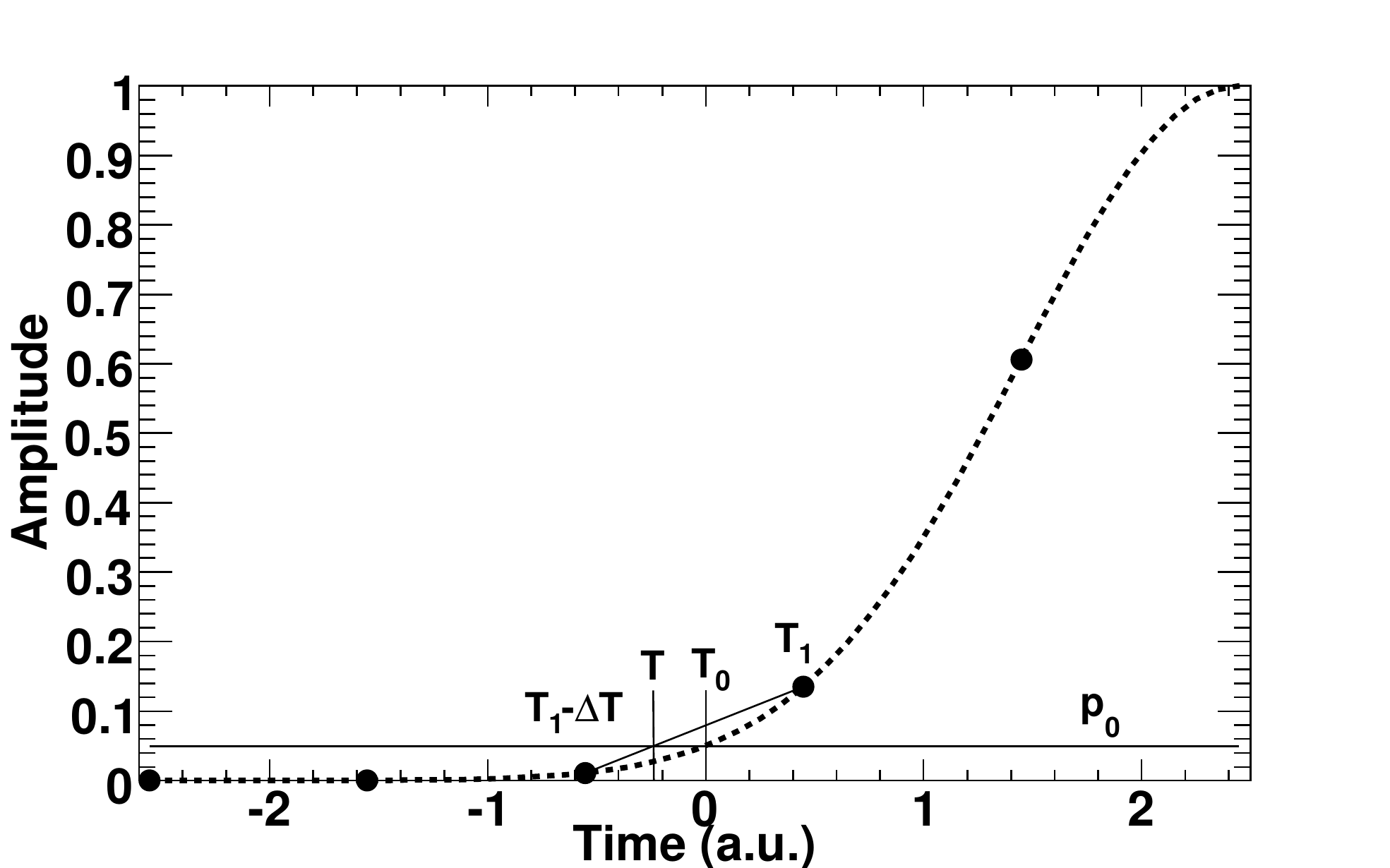}
 \caption[Sampling of a Gaussian function with a time between sampling points equal to the $\sigma$ of the Gaussian function]{Sampling (circles) of a Gaussian function (dashed line) with a time between sampling points, $\Delta T$, equal to the $\sigma$ of the Gaussian function. The threshold, $p_0$, is crossed at a time $T_0$. The first sampling point above the threshold occur at time $T_1$ and the measured time after sampling is $T$.}
 \label{fig:timedef}
\end{figure}

To verify this conclusion one can construct an analytical expression for the measured time distribution due to the finite sampling frequency; see fig.~\ref{fig:timedef} for a definition of the parameters.
The timing measurements are usually carried out by linear interpolation between the sampled points using a straight line. The straight-line equation for $y$ as a function of $x$ is
\begin{equation}
 y = kx+m,
\end{equation}
with the slope
\begin{equation}
 k = \frac{y_2-y_1}{x_2-x_1},
\end{equation}
and the intercept
\begin{equation}
 m = y_2-kx_2.
\end{equation}
From fig.~\ref{fig:timedef} we know that $x_2 = T_1$, $y_2 = f(T_1)$, $x_1 = T_1-\Delta T$ and $y_1 = f(T_1-\Delta T)$. This gives
\begin{equation}
 k = \frac{f(T_1)-f(T_1-\Delta T)}{T_1-T_1+\Delta T} = \frac{f(T_1)-f(T_1-\Delta T)}{\Delta T}
\end{equation} 
and
\begin{equation}
 m = f(T_1)-\frac{f(T_1)-f(T_1-\Delta T)}{\Delta T}T_1.
\end{equation} 
The $y$ value of the interpolated line at a given time $T$ is thus
\begin{equation}
 y = T\frac{f(T_1)-f(T_1-\Delta T)}{\Delta T}+f(T_1)-\frac{f(T_1)-f(T_1-\Delta T)}{\Delta T}T_1.
\end{equation} 
Setting the $y$ value to the threshold $p_0$ gives the time distribution $T(T_{1})$ as
\begin{equation}
 T = T_{1} - \Delta T \frac{f(T_1)-p_{0}}{f(T_1)-f(T_{1}-\Delta T)},\quad\mbox{with $T_0<T_1<T_0+\Delta T$}\label{eq:timedist}.
\end{equation} 

\subsection{Results}

\begin{figure}
 \centering
 \includegraphics[width=\textwidth]{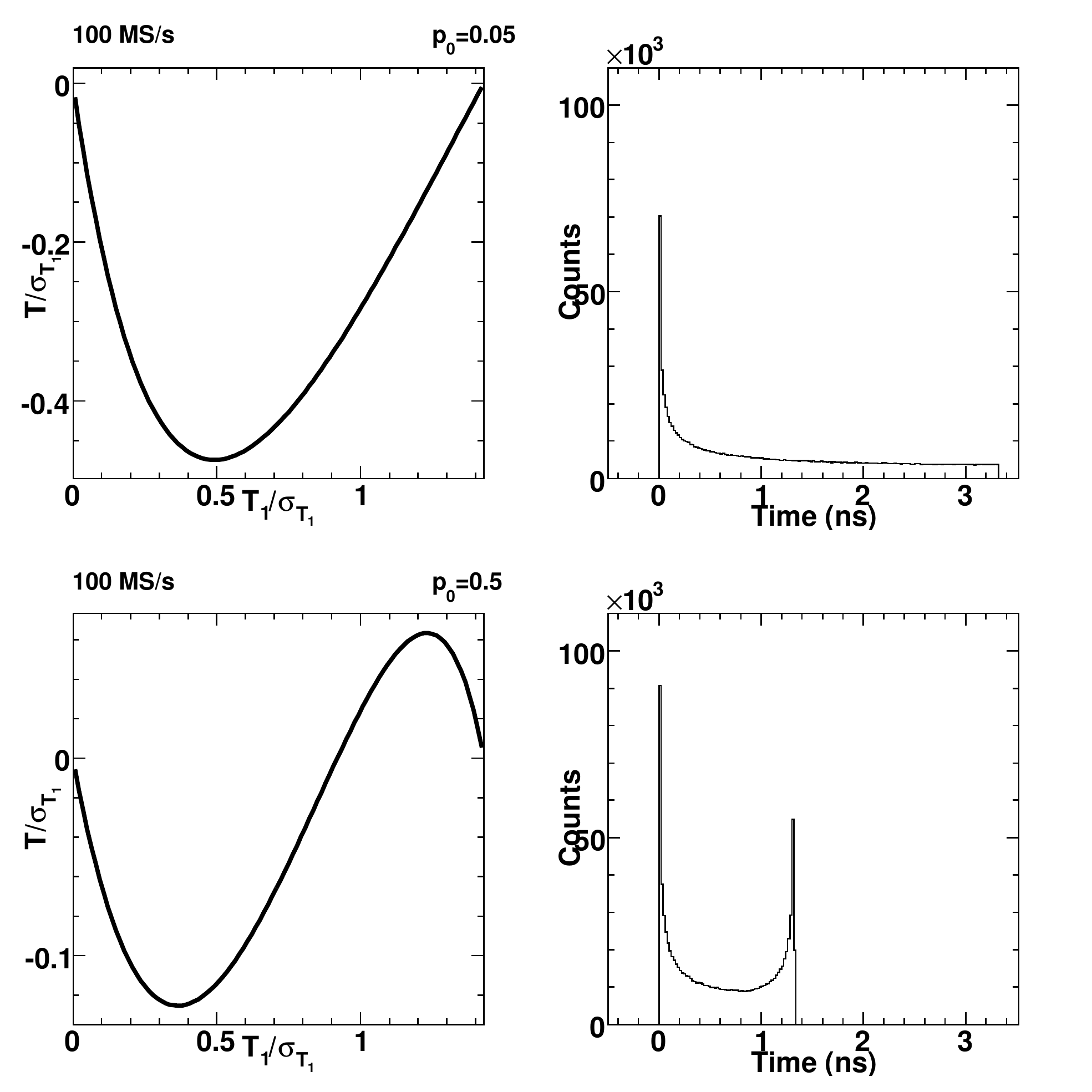}
 \caption[The measured times, $T$, as a function of $T_1$ and the time distributions due to the finite sampling frequency]{In the left panels the measured times, $T$, as a function of $T_1$ are shown for 100~MS/s and thresholds $p_0=0.05$ (upper panel) and $p_0=0.5$ (lower panel). The times are normalized to $\sigma$ of the Gaussian pulse. In the right panels the time distributions due to the finite sampling frequency are shown for the parameters used in the corresponding left panels.}
 \label{fig:allfunctions}
\end{figure}

\begin{figure}
 \centering
 \includegraphics[width=\textwidth]{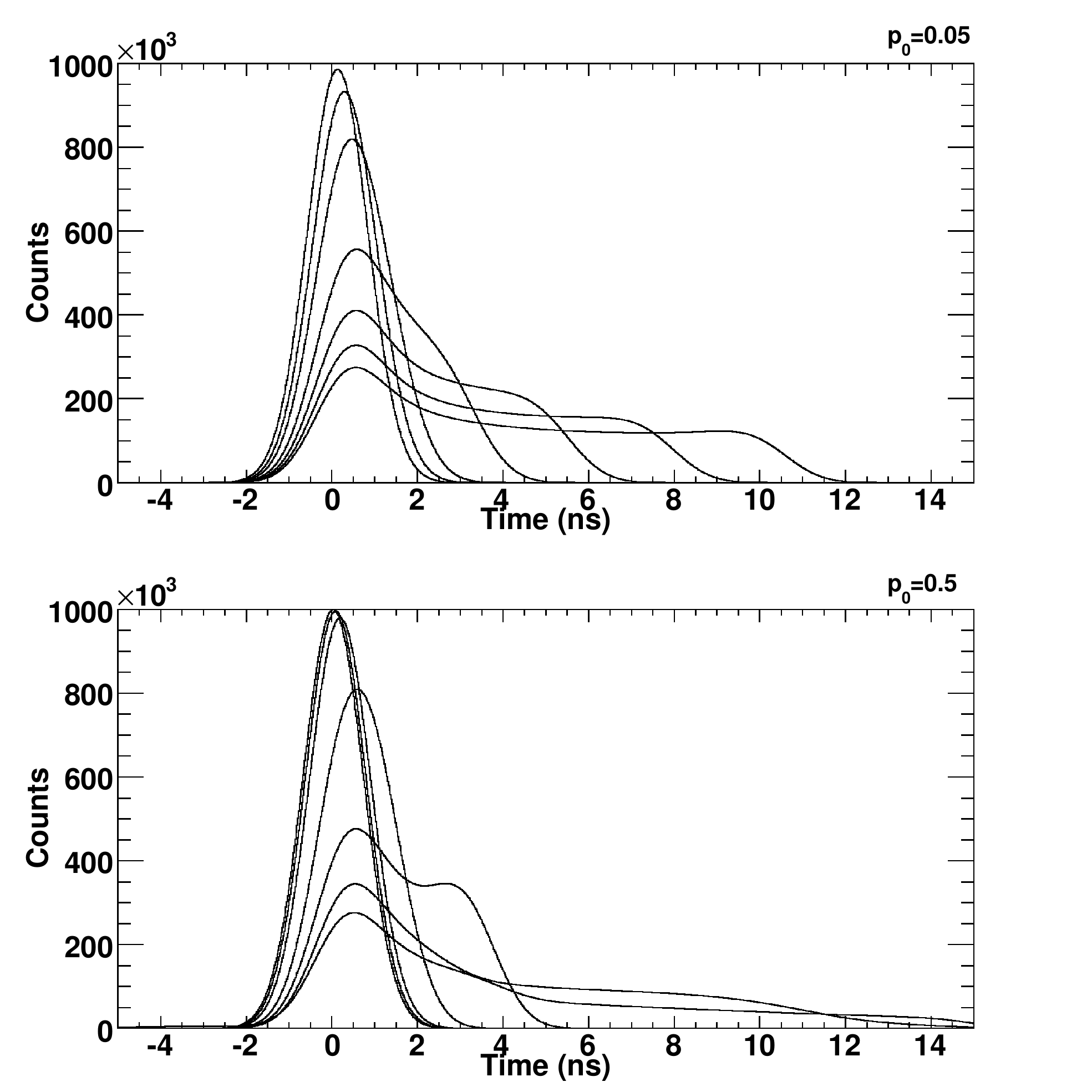}
 \caption[Time distributions folded with a typical Gaussian time resolution of a liquid scintillator detector plus PMT]{Time distributions for $p_0=0.05$ and $p_0=0.5$ folded with a typical Gaussian time resolution of a liquid scintillator detector plus PMT with a time resolution of FWHM=$1.6$~ns. The width of the distributions are decreasing with increasing sampling frequency. The distributions are for the frequencies 300~MS/s, 200~MS/s, 150~MS/s, 100~MS/s, 75~MS/s, 60~MS/s and 50~MS/s. The widest distributions correspond to a sampling frequency of 50~MS/s and the most narrow distributions to a sampling frequency of 300~MS/s.}
 \label{fig:timedists}
\end{figure}

In fig.~\ref{fig:allfunctions} two examples of time distributions, using a Gaussian pulse
\begin{equation}
 f(T_{1}) = \exp{\left(-\frac{T_{1}^{2}}{2 \sigma_{T_{1}}^{2}}\right)},
\end{equation}
with $p_0=0.05$ and $p_0=0.5$ for 100~\ac{MS/s}, are shown.
Assuming that the time between the sampling points and when the pulse passes the threshold is random and uniform one can use eq.~(\ref{eq:timedist}) to generate the time response due to the finite sampling frequency; see fig.~\ref{fig:allfunctions} for two examples. As seen in fig.~\ref{fig:allfunctions}, the finite sampling frequency will result in a non Gaussian time distribution. Folding these time response distributions with a typical intrinsic time resolution of a liquid scintillator detector plus \ac{PMT} of \ac{FWHM}=$1.6$~ns, gives the distributions in fig.~\ref{fig:timedists} for the different sampling frequencies used in ref.~\cite{dpsd}. As can be seen in the figure, the conclusion in ref.~\cite{dpsd}, that 200~\ac{MS/s} sampling frequency is enough for a negligible contribution to the time resolution, is verified for a threshold of $p_0=0.5$.
\cleardoublepage

\chapter{The Neutron Detector Array NEDA\label{sec:neda}}
One of the next generation neutron detector arrays that is being developed is NEDA. NEDA is a part of the SPIRAL2 Preparatory Phase as described in section~\ref{sec:spiral2}. The design of this detector is still in a very early stage. This chapter is intended to be a snapshot of the current status of the project. 

\begin{figure}
 \centering
 \includegraphics[width=0.6\textwidth]{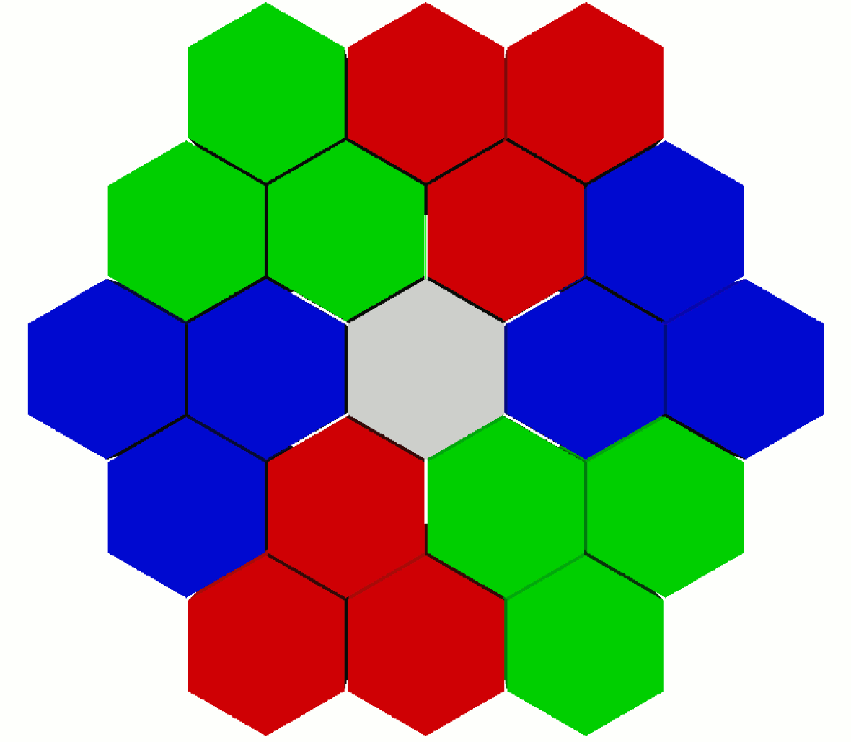}
 \caption[A part of the NEDA array used as a starting point for the simulations]{A part of the NEDA array used as a starting point for the simulations. Courtesy of Tayfun H\"{u}y\"{u}k.}
 \label{fig:neda}
\end{figure}

At the moment a possible NEDA configuration consist of a plane wall of hexagonal detectors elements placed from 0.5~m to $\sim$2~m away from a source of neutrons as shown in fig.~\ref{fig:neda}.
It is very important to carefully study different geometries both to maximize efficiency and to minimize cross talk due to multiple scattering of neutrons between the detectors. 
The cross talk is due to the effect that neutrons deposit energy mainly by elastic scattering with the protons in the liquid. The energy of the recoil protons can have values from zero up to the incoming neutron energy, depending on the scattering angle. Because of this there exists a probability that the neutron will scatter into a neighboring detector and interact again, causing severe errors in the counting of the number detected neutrons, which is used to measure the number of neutrons emitted in the reaction. Several methods attempting to correct for this exist \cite{2004NIMPA.528..741,1997NIMPA.385..166C}. For example a method based on the \ac{TOF} difference between the different segments has shown to give good results \cite{2004NIMPA.528..741} when the segments are sufficiently far away while neighboring segments still are problematic. This is important to take into account when designing the geometry of a new array.

The main plan is that NEDA will be filled with the same liquid scintillator, BC-501A, as is commonly used in other neutron detector arrays. Studies are also being made regarding  the possibility to use other detector materials, for example BC-537, which is a deuterated liquid scintillator with a deuteron to proton ratio of 114:1. This liquid is being evaluated for the DESCANT project \cite{descant} and it may give some additional energy resolution and cross-talk rejection properties.

\section{Readout\label{ss:readout}}

One way to improve the efficiency of the neutron detector array is to improve the efficiency of the readout of the scintillation light. As mentioned in section~\ref{ss:detectors}, the standard way to read out a liquid scintillator is to couple the liquid to a \ac{PMT}. The \ac{PMT} typically consist of a photocathode of a bialkali metal alloy, like Sb-Rb-Cs or Sb-K-Cs \cite{sommer}, a series of dynodes to multiply the photoelectrons and an anode to read out the signal. Instead of using the standard setup with a photocatode and a series of dynodes one can also use a hybrid setup with a photocatode that is multiplied by an \ac{APD}. Using a hybrid setup one will have the opportunity to have a position sensitive \ac{PMT}. A technology for that is for example used in the LHCb experiment \cite{LHCb}. The bialkali photocathode has a quantum efficiency of maximum $\sim25$\% and a sensitivity well matched to the most common scintillator materials. This is one of the reasons why it is so frequently used for reading out scintillation counters. If it is possible to increase this value, the performance of the neutron detector array will be much higher, something that is critical in order to study rare reaction channels. A useful property of \ac{PMT}s is, however, that regular \ac{PMT}s are available in a large variety of sizes from a diameter of a few mm up to about 50~cm diameter.

One alternative to a \ac{PMT} for readout is to use either a regular photodiode or an \ac{APD}. Today regular photodiodes can be manufactured up to sizes of about $20 \times 20$~mm$^2$ while \ac{APD}s are limited to about $10 \times 10$~mm$^2$. The quantum efficiency of the photodiodes can be very high, up to 85\% at the peak of maximum sensitivity, but the spectral response of this kind of diodes usually shows a very sharp drop at lower wavelengths. This is due to the fabrication of the photodiode where a thin film of 80-100~nm p-doped material is implanted on the silicon wafer, which makes it difficult to get high sensitivity at shorter wavelengths. There are, however, also some examples of large-area short-wavelength \ac{APD}s, like the Hamamatsu S8664 which is planned to be used for the readout for the $\bar{\textrm{P}}$ANDA electromagnetic calorimeter \cite{panda_emc_tdr}. Regular and avalanche photodiodes can be useful for feasibility tests but the size and wavelength response are issues that need to be resolved in order for photodiodes to be used by NEDA.

Silicon \ac{PMT}s are another kind of position sensitive readout that has gained popularity in some fields lately \cite{Buzhan:2001xq}. These consist of an array of around 1000 independent \ac{APD}s per mm$^{2}$. The quantum efficiency and gain of this kind of readout is quite similar to regular \ac{PMT}s. The main advantages are good time resolution, no dependency on external magnetic fields and an extremely compact design. Although these parameters are of importance as well there must still be a substantial development regarding quantum efficiency and gain for them to be considered for NEDA.

Another very promising technology is the ultra-pure bialkali photocathode \ac{PMT}s that have been made available quite recently by Hamamatsu. By using a fine tuned deposition process one could achieve ultra-pure photocathode materials and reach a quantum efficiency of up to 43\% \cite{bialkali_nim}. These \ac{PMT}s are currently available in sizes of up to 76~mm diameter, which is probably not enough for NEDA. If larger sizes are developed they might be a very promising candidate for replacement of the regular \ac{PMT}s. More tests of both \ac{APD}s and ultra-pure bialkali photocathode \ac{PMT}s would give valuable information for the design of the development of neutron detectors for the next generation of radioactive ion-beam facilities.
\cleardoublepage

\chapter{Summary and Outlook\label{sec:summary}}
In this work the performance of the Neutron Wall detector array and digital \acl{PSD} of neutrons and $\gamma$ rays has been studied. The detection process in liquid scintillators has been discussed in detail. The physics motivation for a new neutron detector array has also been briefly discussed and a snapshot of the development of NEDA has been presented.

To increase the efficiency of the Neutron Wall it is suggested to increase the \acl{HV} of the \acl{PMT}s. The apparent decrease in quality of the separation in the one-dimensional \acl{ZCO} projections due to low-energy neutron interactions could be corrected for if the \acl{ZCO} is used together with the \acl{QVC} and \acl{TOF}. For the next generation neutron detector array, NEDA, other readout systems have been discussed. To further investigate alternative read out, the scintillator BC-501A should be tested together with one of the $\bar{\textrm{P}}$ANDA \acl{APD}s, especially regarding \acl{PSD} properties of this kind of detector system.

The other main part of this work has been to develop digital \acl{PSD} algorithms and test the performance of these with respect to sampling frequency and bit resolution of the \acs{ADC}. The results from these measurements show that an \acl{ADC} with a bit resolution of 12 bits and a
sampling frequency of 100 \acs{MS/s} is adequate for \acl{PSD} of neutrons and ${\gamma}$ rays for a neutron energy range of
0.3--12 MeV. One limitation of these tests has been the available \acs{ADC}, in this work 300 \acs{MS/s} and 14 bits. In order to try different combinations of sampling frequency and bit resolution it is desirable to simulate realistic pulse shapes and the sampling of these. As input to these simulations, it is necessary to measure the time constants of the decay components, as well as to measure the relative amount of the different components for different particle species. This is a project that is still ongoing \cite{paulas}. It has also been shown that 200~\acs{MS/s} sampling frequency is enough for a negligible contribution to the time resolution.
\cleardoublepage

\chapter*{Acknowledgements}
\addcontentsline{toc}{chapter}{Acknowledgements}
\ackgroup{First of all I would like to thank my supervisors Johan Nyberg and Ayse Ata\c{c} for all their help during this time. I would also like to thank my supervisors from my Masters thesis, Stephan Pomp and Jan Blomgren, for showing me the joy of experimental nuclear physics, thus keeping me on the right track.}

\ackgroup{It is very exciting to work in the NEDA collaboration and see all different ideas merging together into the next neutron detector array. Thanks to Jose Javier Valiente Dobon  for organizing this and the rest of the NEDA collaboration for their efforts.}

\ackgroup{The work on the Neutron Wall was very rewarding much thanks to Marcin Palacz, Gilles de France and Paula Salvador Casti\~{n}eira. The work, dinners, tourism and company during the SPIRAL2 Week were all very nice.}

\ackgroup{I would also like to acknowledge the people in the Ph.D. student councils, both at the faculty and at the student union, for all energy they put in keeping Ph.D. studies in Uppsala a well spent time for everyone.}

\ackgroup{I am also grateful to Sophie Grape for reading the manuscript, giving valuable comments and, just when I was finished, forcing me to rearrange large parts of this work.}

\ackgroup{I would also like to acknowledge all fellow Ph.D. students at the department that keeps everyday life joyful: Karin Sch\"{o}nning; Erik Thom\'{e}, for company during obscure seminars and for allowing me to use his chair now and then; Sophie Grape, for discussions, cakes, food, tea, gym, advice, gossip, movies and everything else; Bengt S\"{o}derbergh; Carl-Oscar Gullstr\"{o}m; Patrik Adlarson; Henrik Petr\'{e}n; Kristofer Jakobsson, for being a good office mate; Mikael H\"{o}\"{o}k, for every zombie movie session throughout the years --- keep on rotting in a free world!; David Eriksson; Oscar St\aa{}l; Martino Olivo; Olle Engdeg\aa{}rd; Camille B\'{e}langer-Champagne, for gym company and everything else; Martin Flechl; Elias Coniavitis, for doing his best keeping up the Thursday spirit; and Andrea Palaia. And all Ph.D. and masters students that have come and gone during these two years, and that will come and go during the coming two years. And Enzo, although you are not a Ph.D. student you are still OK.}

\ackgroup{There are also other friends that should be mentioned: Karin and Jenny, just for being very good friends; Freddyboy, for filling my life with music; Klas-Herman and Hanna Victoria M\"{o}rck, for playing board games, giving me coffee and getting me involved in more things than what is probably good for me; Henke and Carolina, for being the wonderful persons you are; and Martin, Lasse, Thomas and the other old friends from Hudiksvall that always make visiting fun.}

\ackgroup{Jag vill ocks\aa{} tacka Leena som funnits d\"{a}r hela mitt liv och Tage som funnits d\"{a}r n\"{a}stan hela mitt liv, och som b\aa{}da kommer att forts\"{a}tta finnas hos mig resten av livet.}

\ackgroup{Thanks to everyone!}

\begin{flushright}
P\"{a}r-Anders S\"{o}derstr\"{o}m\\
\includegraphics{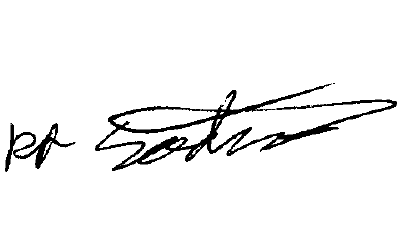}
\end{flushright}
\cleardoublepage

\addcontentsline{toc}{chapter}{References}
\bibliographystyle{pa}

\end{document}